\documentclass[final]{siamltex}



\usepackage{amsmath,amssymb,amsthm}

\usepackage{graphicx}
\usepackage{subcaption}
\usepackage{color}
\usepackage{url}
\usepackage{cases}
\usepackage{algorithmic}
\usepackage{algorithm}
\usepackage[normalem]{ulem}
\usepackage{tikz}
\usetikzlibrary{shapes.geometric, arrows}

\title{An efficient and accurate MPI-based parallel simulator for streamer discharges in three dimensions}

\author{Bo Lin\thanks{Department of Mathematics, National University of Singapore, 10 Lower Kent Ridge Road, Singapore, 119076.({\tt linbo@u.nus.edu})}
	\and Chijie Zhuang\thanks{State Key Lab of Power Systems and Department of Electrical Engineering, Tsinghua University, Beijing 100084, China.({\tt chijie@tsinghua.edu.cn, corresponding author})}
	\and Zhenning Cai\thanks{Department of Mathematics, National University of Singapore, 10 Lower Kent Ridge Road, Singapore, 119076. ({\tt matcz@nus.edu.sg}) }
	\and Rong Zeng\thanks{State Key Lab of Power Systems and Department of Electrical Engineering, Tsinghua University, Beijing 100084, China.({\tt zengrong@tsinghua.edu.cn})}
	\and Weizhu Bao\thanks{Department of Mathematics, National University of Singapore, 10 Lower Kent Ridge Road, Singapore, 119076. ({\tt matbaowz@nus.edu.sg})}
}

\newcommand\deletei{\bgroup\markoverwith{\textcolor{red}{\rule[0.5ex]{2pt}{2pt}}}\ULon}
\newcommand\deleteii{\bgroup\markoverwith{\textcolor{blue}{\rule[0.5ex]{2pt}{2pt}}}\ULon}

\begin{document}

\maketitle

\begin{abstract}
In this paper, we propose an efficient and accurate message-passing interface (MPI)-based parallel simulator for streamer discharges in three dimensions using the fluid model. First, we propose a new second-order semi-implicit scheme for the temporal discretization of the model that relaxes the dielectric relaxation time restriction. Moreover, it requires solving the Poisson-type equation only once at each time step, while the classical second-order explicit scheme typically need to do twice. Second, we introduce a geometric multigrid preconditioned FGMRES solver that dramatically improves the efficiency of solving the Poisson-type equation with either constant or variable coefficients. We show numerically that no more than 4 iterations are required for the Poisson solver to converge to a relative residual of 10$^{-8}$ during streamer simulations; the FGMRES solver is much faster than R\&B SOR and other Krylov subspace solvers. Last but not least, all the methods are implemented using MPI. The parallel efficiency of the code and the fast algorithmic performances are demonstrated by a series of numerical experiments using up to 2560 cores on the Tianhe2-JK clusters. For applications, we study a double-headed streamer discharge as well as the interaction between two streamers, using up to 10.7 billion mesh cells.
\end{abstract}


\pagestyle{myheadings}
\thispagestyle{plain}
\markboth{Bo Lin and Chijie Zhuang et al.}{An efficient and accurate MPI based parallel simulator for streamer discharges in three dimensions}

\section{Introduction}\label{Introduction}

A streamer is a cold plasma that is common in both nature and industrial processes. As the building blocks of long air-gap discharges, streamer discharges are associated with many insulation problems \cite{yuesheng} such as flashovers along an insulator, air-gap breakdowns in DC power systems \cite{she}, or lightning bolts \cite{hve} where the sprites triggered by the strong quasi-electrostatic field generated by intense cloud-to-ground lightning flashes have been found to be filament streamer discharges \cite{nc}. 

In this paper, we focus on the numerical methods for streamer discharge simulations in three dimensions (3D), therefore we use a minimal model that incorporates the essential mechanism of the phenomena \cite{ebert}. The minimal three-dimensional model for simulating streamer discharges consists of two convection-dominated transport equations coupled with a Poisson equation for the electrical potential and the field:
\begin{align}
\left\{ \begin{aligned}
& \frac{\partial n_e}{\partial t} - \nabla \cdot (\mu_e \vec{E} n_e) - \nabla \cdot (D_e \nabla n_e) = \alpha(|\vec{E}|) \mu_e | \vec{E} | n_e , \\
& \frac{\partial n_p}{\partial t} + \nabla \cdot (\mu_{p} \vec{E} n_p) = \alpha(|\vec{E}|) \mu_{e} | \vec{E} | n_e, \qquad \qquad \qquad \vec{x} \in \Omega, \ \  t>0, \\
& - \Delta \phi = \frac{e}{\varepsilon_0} (n_p-n_e), \qquad \vec{E} = -\nabla \phi,
\end{aligned}
\right.
\label{eM}
\end{align}
where $\vec{x}=(x,y,z)^T$ and $\Omega \subset \mathbb{R}^3$ is a bounded domain; $n_e=n_e(\vec{x},t)$ and $n_p=n_p(\vec{x},t)$ denote the densities of electrons and positive ions, respectively; $\phi=\phi(\vec{x},t)$ and $\vec{E}=\vec{E}(\vec{x},t)$ denote the electric potential and electric field, respectively;  $\mu_e$ and $\mu_p$ ($\mu_e>0$, $\mu_p>0$) are mobility constants for electrons and positive ions, respectively; $D_e$ is a diagonal matrix $D_e =\diag(D_{e,x},D_{e,y},D_{e,z})$, and $D_{e,x}$, $D_{e,y}$ and $D_{e,z}$ are the diffusion coefficients in $x$, $y$, $z$ directions, respectively. Here, $\alpha=\alpha(|\vec{E}|)$ is the effective ionization coefficient, and the parameters $e$ and $\varepsilon_0$ are the elementary charge and the vacuum dielectric permittivity, respectively. 

To model the streamer discharge between two parallel plates, a cubic domain $\Omega=(x_0,x_1) \times (y_0,y_1) \times (z_0, z_1)$ is considered. Dirichlet boundary conditions are applied for the potential, $\phi$, on the upper and lower plate electrodes (i.e., $\phi|_{z=z_1}=\phi_0$ where $\phi_0$ is a constant denotes applied potential, and $\phi|_{z=z_0}=0$); and homogeneous Neumann boundary conditions, which are $\frac{\partial \phi}{\partial x} |_{x=x_0,x_1}=0$ and $\frac{\partial \phi}{\partial y} |_{y=y_0,y_1}=0$, are applied on other four sides. Initial conditions for $n_e$ and $n_p$ are given as
\begin{equation}
n_e(\vec{x},t=0)=n_{e,0}(\vec{x}), \qquad n_p(\vec{x},t=0)=n_{p,0}(\vec{x}), \qquad \vec{x} \in \bar{\Omega}.
\label{initial}
\end{equation}
For simplicity, the plasma is initially assumed to be electrically neutral everywhere, which gives  $n_{e,0}(\vec{x}) = n_{p,0}(\vec{x}) =\tilde{n}(\vec{x})$ with $\tilde{n}(\vec{x})$ a given function. Homogeneous Neumann boundary conditions are applied at all the boundaries for $n_e$, and at all inflow boundaries for $n_p$.

Continuous efforts have been made to simulate streamer discharges over the past few decades. In the 1980s and 1990s, the flux-corrected transport(FCT) technique \cite{book,zalezak} was widely used. It was combined with the finite difference method (FDM) and finite element method (FEM) to overcome the numerical oscillations that occur when classical linear schemes are used to solve convection-dominated equations \cite{morrow2, morrow1}. Later, the finite volume method (FVM) became popular due to the property of local conservation \cite{ebert,ZAKARI2015473}. Motivated by the successes of FVM and FEM, the discontinuous Galerkin (DG) method, which uses a finite element discretization with discontinuous basis functions and incorporates the ideas of numerical fluxes and slope limiters from the high-resolution FDM and FVM, was used to simulate the streamers \cite{cicpdg,ldg,2ddg}. These improvements in the numerical methods achieved great progress in streamer simulations \cite{sixcode}, especially in two-dimensional (2D) cases where the streamer is assumed to be axisymmetric. However, compared with 2D simulations, studies of real three-dimensional simulations are much fewer, and are mostly done by limited groups \cite{marskar2019adaptive, 0022-3727-51-9-095206, liu, teunissen2017, TEUNISSEN2018156, VILLA2015369,   VILLA2017233, VILLA2017687}. 

The difficulty in three-dimensional simulations lies in the fine meshes needed to simulate rapid variations in the solution. Streamer discharges propagate at dramatic speeds, e.g., at $10^6$ m/s, as shown in Fig. 7 of \cite{speed}. During this rapid transient process, the electric field in the discharge channel, which is one of the key parameters dominating the development of a streamer, varies significantly both temporally and spatially. After streamer inception, the electric field at its head is greatly enhanced due to the net charge accumulation, which further accelerates the ionization and charge accumulation. Thus, a sharp charge density profile forms at the streamer's head. Capturing the structures of the charge carriers in a simulation requires a very high-resolution spatial grid. Typically, the order of magnitude for the grid size adopted in previous simulations has been characterized by micrometres \cite{movingmesh2007, teunissen2017}, which is tiny compared with the characteristic length of the problem at the scale of, e.g., centimetres. Consequently, the maximal allowed time step is restricted to the order of several picoseconds or even smaller when explicit schemes are used. In addition, because the Poisson equation and transport equations for the charge carriers are coupled together, the time step is further restricted by the dielectric relaxation time, i.e., $\varepsilon_0/e \max(\mu_p n_p + \mu_e n_e)$, which is also typically on the order of several picoseconds. For these reasons, even two-dimensional simulations require long computational times, let alone three-dimensional simulations which have thousands times the number of degrees of freedom (DOFs) in 2D simulations, for even a small domain in 3D. Thus, it seems parallel computing is the only possible way to efficiently construct large-scale three-dimensional simulations for streamer discharges.

Recently, Teunissen and Ebert reported on 3D streamer simulations using the parallel adaptive Afivo framework \cite{teunissen2017}, which features adaptive mesh refinement (AMR), geometric multigrid methods for the Poisson equation and OpenMP parallelism. AMR performs well, however, further improvement could be made by replacing the OpenMP parallelism with message-passing interface (MPI) libraries to fully exploit the power of clusters, especially for simulations of very long streamers. Another advance in the MPI-based simulation was reported by Plewa, Eichwald, and Ducasse et al. \cite{0022-3727-51-9-095206}, who used the successive over-relaxation iterative solver in the red and black strategy (R\&B SOR) as the Poisson solver, and tested the parallelization and the scalability with cell numbers ranging from 8–512 million and core numbers from 20–1600. Their use of high-performance computing clusters with an MPI implementation reduced the computational time. 

These previous works suggest an efficient simulator has advantages in the following aspects: (1) an efficient time integration scheme, which may reduce the time marching steps; (2) a fast algebraic elliptic solver, which accelerates the solution of the Poisson equation that dominates the total computing time; (3) a good parallelization, which allows to utilize the full power of modern clusters; (4) adaptive mesh strategies, which can reduce the number of DOFs. This paper contributes to the first three aspects.
First, we propose a new second-order semi-implicit scheme for temporal discretization. In particular, the scheme is stable when the time step exceeds the dielectric relaxation time. We numerically demonstrate that it is second-order accurate in time. Moreover, at each time step, it requires solving the Poisson equation only once, while previous second-order schemes typically need twice \cite{teunissen2017, cicpdg}. Note that solving the Poisson equation is the most expensive calculation in the simulator. Second, we adopt the geometric multigrid preconditioned FGMRES solver with Chebyshev iteration as the smoother in the multigrid preconditioner, which dramatically improves the efficiency of solving Poisson equations with either constant or variable coefficients. We show numerically that the multigrid preconditioned FGMRES algorithm is more efficient than other Krylov subspace based methods and R\&B SOR. We implement all methods using MPI, and the code runs with good parallel efficiency on the Tianhe2-JK cluster using more than 2500 cores. The numerical experiments demonstrate the good performance of the algorithms. Finally, we study a double-headed streamer discharge as well as the interaction of two streamers using up to 10.7 billion mesh cells.

This paper is organized as follows. In Section 2, after reviewing some existing temporal discretization schemes, we describe our new second-order semi-implicit temporal scheme in details and briefly show the spatial discretization. Multigrid preconditioned FGMRES elliptic solver is described in Section 3, and the MPI parallelism is briefly described in Section 4. In Section 5, we use a one-dimensional dimensionless example to illustrate the convergence order and stability of different temporal schemes, and then take 3D examples to show the scalability of the parallel implementation and the performance of different algebraic elliptic solvers. Section 6 gives simulation results of a double-headed streamer propagation as well as the interaction between two streamers. Conclusions are drawn in Section 7.

\section{Numerical discretization} \label{NumericalMethod}
In this section, we first focus on temporal discretization and present a new second-order semi-implicit scheme. Then, we introduce the finite volume method for spatial discretization.
\subsection{A second-order semi-implicit temporal discretization} \label{2nd_order}
Let $t_0 = 0$, $\tau_n>0$ be the time step at $n$-th step, and $t_{n+1} = t_n + \tau_n$ for $n \geq 0$. We use $n_e^n=n_e^n(\vec{x})$, $n_p^n=n_p^n(\vec{x})$, $\phi^n = \phi(\vec{x})$ and $\vec{E}^n=\vec{E}^n(\vec{x})$ to denote the associated quantities to be approximated at time $t_n$. To avoid solving nonlinear algebraic equations, explicit schemes are frequently used for time discretization, among which the forward Euler scheme is used to discretize the model \eqref{eM} as
\begin{align}
\left\{ \begin{aligned}
& \frac{n_e^{n+1}-n_e^n}{\tau_n} - \nabla \cdot (\mu_e \vec{E}^n n_e^n) - \nabla \cdot (D_e \nabla n_e^n) = \alpha(|\vec{E}^n|) \mu_e | \vec{E}^n | n_e^n , \\
& \frac{n_p^{n+1}-n_p^n}{\tau_n} + \nabla \cdot (\mu_{p} \vec{E}^n n_p^n) = \alpha(|\vec{E}^n|) \mu_{e} | \vec{E}^n | n_e^n, \qquad \qquad \qquad \vec{x} \in \Omega. \\
& - \Delta \phi^n = \frac{e}{\varepsilon_0} (n_p^n-n_e^n), \qquad \vec{E}^n = -\nabla \phi^n,
\end{aligned}
\right.
\label{explicit}
\end{align}
At each time step, the potential $\phi^n$ is first calculated by the Poisson equation, and then $n_e^{n+1}$ and $n_p^{n+1}$ are obtained subsequently. The Poisson equation need to be solved once at each time step. 

It is easy to see the scheme (\ref{explicit}) is only first order in time, and it has been upgraded to second order by Heun's method, as is used in \cite{teunissen2017}. The first stage of Heun's method is to solve $\phi^n$, $n_e^*$ and $n_p^*$ from $n_e^n$ and $n_p^n$,
\begin{align}
\left\{ \begin{aligned}
& \frac{n_e^{*}-n_e^n}{\tau_n} - \nabla \cdot (\mu_e \vec{E}^n n_e^n) - \nabla \cdot (D_e \nabla n_e^n) = \alpha(|\vec{E}^{n}|) \mu_e | \vec{E}^n | n_e^n , \\
& \frac{n_p^{*}-n_p^n}{\tau_n} + \nabla \cdot (\mu_{p} \vec{E}^n n_p^n) = \alpha(|\vec{E}^{n}|) \mu_{e} | \vec{E}^n | n_e^n, \qquad \qquad \qquad \vec{x} \in \Omega, \\
& - \Delta \phi^n = \frac{e}{\varepsilon_0} (n_p^n-n_e^n), \qquad \vec{E}^n = -\nabla \phi^n, \\
\end{aligned}
\right.
\label{heun_ex_1}
\end{align}
and then evolve the solution through one more stage to obtain $n_e^{**}$ and $n_p^{**}$:
\begin{align}
\left\{ \begin{aligned}
& \frac{n_e^{**}-n_e^*}{\tau_n} - \nabla \cdot (\mu_e \vec{E}^* n_e^*) - \nabla \cdot (D_e \nabla n_e^*) = \alpha(|\vec{E}^{*}|) \mu_e | \vec{E}^* | n_e^* , \\
& \frac{n_p^{**}-n_p^*}{\tau_n} + \nabla \cdot (\mu_{p} \vec{E}^* n_p^*) = \alpha(|\vec{E}^{*}|) \mu_{e} | \vec{E}^* | n_e^*, \qquad \qquad \qquad \vec{x} \in \Omega. \\
& - \Delta \phi^* = \frac{e}{\varepsilon_0} (n_p^*-n_e^*), \qquad \vec{E}^* = -\nabla \phi^*, \\
\end{aligned}
\right.
\label{heun_ex_2}
\end{align}
The final solution at $t_{n+1}$ is constructed by
\begin{align}
n_e^{n+1} = \frac{1}{2} \left( n_e^n + n_e^{**} \right), \qquad n_p^{n+1} = \frac{1}{2} \left( n_p^n + n_p^{**} \right), \qquad \vec{x} \in \Omega.
\label{heun_ex_3}
\end{align}
This temporal scheme possesses a second-order time accuracy and has been used (e.g., in \cite{movingmesh2007,ebert}).

We emphasize that the second-order explicit scheme shown in (\ref{heun_ex_1})--(\ref{heun_ex_3}) requires solving the Poisson equation twice at each time step (from $t_n$ to $t_{n+1}$). Moreover, it was suggested in \cite{diel1987, teunissen2017} that these explicit schemes need to satisfy the dielectric relaxation time constraint, i.e.,
\begin{equation} \label{drt}
\tau_n \leq \frac{\varepsilon_0}{e \max(\mu_p n_p^n +\mu_e n_e^n)}, \qquad n \geq 0.
\end{equation}

To relax this time constraint \eqref{drt}, semi-implicit schemes were introduced \cite{ventzek1993two,villa2013asymptotic}. In \cite{villa2013asymptotic}, Villa et al. proposed a first-order semi-implicit scheme with a rigorous asymptotic preserving property. Here we present the scheme with a slight modification as
\begin{align}
\left\{ \begin{aligned}
& \frac{n_e^{n+1}-n_e^n}{\tau_n} - \nabla \cdot (\mu_e \vec{E}^{n+1} n_e^n) - \nabla \cdot (D_e \nabla n_e^n) = \alpha(|\vec{E}^{n+1}|) \mu_e | \vec{E}^{n+1} | n_e^{n} ,\\
& \frac{n_p^{n+1}-n_p^n}{\tau_n} + \nabla \cdot (\mu_{p} \vec{E}^{n+1} n_p^n) = \alpha(|\vec{E}^{n+1}|) \mu_{e} | \vec{E}^{n+1} | n_e^{n}, \qquad \qquad \qquad \vec{x} \in \Omega.\\
& - \Delta \phi^{n+1} = \frac{e}{\varepsilon_0} (n_p^{n+1}-n_e^{n+1}), \qquad \vec{E}^{n+1} = -\nabla \phi^{n+1},
\end{aligned}
\right.
\label{semi-implicit}
\end{align}
In \cite{villa2013asymptotic}, a fully implicit source term $\alpha(|\vec{E}^{n+1}|) \mu_{e} | \vec{E}^{n+1} |  n_e^{n+1}$ was adopted; however, our simplification of the source term in \eqref{semi-implicit} does not affect the proof of the asymptotic preserving property. A comparison of \eqref{semi-implicit} and \eqref{explicit} shows that the main difference between the semi-implicit scheme and the explicit schemes lies in whether the electric field is treated implicitly. As demonstrated in \cite{villa2013asymptotic}, when the reference states of $n_e$, $n_p$ and $\vec{E}$ are bounded, the time step $\tau_n$ is no longer restricted by the dielectric relaxation time.

Although \eqref{semi-implicit} is a semi-implicit discretization of \eqref{eM}, thanks to the structure of \eqref{semi-implicit}, it can be solved explicitly by rewriting the Poisson equation as a variable coefficient elliptic equation or Poisson-type equation \cite{villa2013asymptotic}. A subtraction of the first two equations in \eqref{semi-implicit} yields
\begin{align}
\frac{(n_p^{n+1}-n_e^{n+1})-(n_p^n -n_e^n)}{\tau_n} + \nabla \cdot (\mu_{p} \vec{E}^{n+1} n_p^n) + \nabla \cdot (\mu_e \vec{E}^{n+1} n_e^n) + \nabla \cdot (D_e \nabla n_e^n) = 0.
\label{subtract}
\end{align}
Then, we plug the expression of $(n_p^{n+1}-n_e^{n+1})$ in \eqref{subtract} into the Poisson equation in \eqref{semi-implicit}, and obtain an elliptic equation:
\begin{align}
-\nabla \cdot \left( \left( \frac{\varepsilon_0}{e} +\tau_n \left( \mu_p n_p^n + \mu_e n_e^n \right) \right) \nabla \phi^{n+1}  \right) = n_p^n - n_e^n -\tau_n \nabla \cdot(D_e\nabla n_e^n ).
\label{e15}
\end{align}
After solving the variable coefficient elliptic problem \eqref{e15}, we obtain $\phi^{n+1}$. Then, we can calculate $\vec{E}^{n+1}=-\nabla \phi^{n+1}$, and evolve the first two equations in \eqref{semi-implicit} to obtain $n_e^{n+1}$ and $n_p^{n+1}$. 

Scheme (\ref{semi-implicit}) is only first order accurate in time, which will be numerically demonstrated later in Table \ref{t6}. Here we propose a new second-order semi-implicit scheme for (\ref{eM}). Our scheme can be regarded as a predictor-corrector method. First, we calculate a prediction $n_e^{n+1/2}$, $n_p^{n+1/2}$, $\phi^{n+1/2}$ and $\vec{E}^{n+1/2}$ at time $t_n+\tau_n/2$ using the first-order semi-implicit scheme (\ref{semi-implicit}), i.e.,
\begin{align}
\left\{
\begin{aligned}
& \frac{n_e^{n+1/2}-n_e^n}{\tau_n/2} - \nabla \cdot (\mu_e \vec{E}^{n+1/2} n_e^{n}) - \nabla \cdot (D_e \nabla n_e^{n}) = \alpha(|\vec{E}^{n+1/2}|)\mu_e |\vec{E}^{n+1/2}| n_e^n, \\
& \frac{n_p^{n+1/2}-n_p^n}{\tau_n /2} + \nabla \cdot (\mu_p \vec{E}^{n+1/2}n_p^n) = \alpha(|\vec{E}^{n+1/2}|)\mu_e |\vec{E}^{n+1/2}| n_e^n, \label{e1} \qquad \qquad \qquad \vec{x} \in \Omega.\\
& - \Delta \phi^{n+1/2} =\frac{e}{\varepsilon_0} ( n_p^{n+1/2} - n_e^{n+1/2} ), \qquad \vec{E}^{n+1/2} = -\nabla \phi^{n+1/2}, 
\end{aligned}
\right.
\end{align}
Then, we get a correction of $n_e$ and $n_p$ using a midpoint scheme, which yields \eqref{e2}
\begin{align}
\left\{
\begin{aligned}
& \frac{n_e^{n+1}-n_e^n}{\tau_n} - \nabla \cdot (\mu_e \vec{E}^{n+1/2} n_e^{n+1/2}) - \nabla \cdot (D_e \nabla n_e^{n+1/2}) = \alpha(|\vec{E}^{n+1/2}|) \mu_e |\vec{E}^{n+1/2}| n_e^{n+1/2}, \label{e2}\\
& \frac{n_p^{n+1}-n_p^n}{\tau_n} + \nabla \cdot (\mu_p \vec{E}^{n+1/2}n_p^{n+1/2}) = \alpha(|\vec{E}^{n+1/2}|) \mu_e |\vec{E}^{n+1/2}| n_e^{n+1/2},
\end{aligned} \vec{x} \in \Omega.
\right.
\end{align}
The potential $\phi^{n+1/2}$ and electric field $\vec{E}^{n+1/2}$ are already predicted at time $t_n+\tau_n/2$ by solving the following variable coefficient elliptic equation derived from \eqref{e1}:
\begin{align}
-\nabla \cdot \left( \left( \frac{\varepsilon_0}{e} + \frac{\tau_n}{2} \left( \mu_p n_p^n + \mu_e n_e^n \right) \right) \nabla \phi^{n+1/2}  \right) = n_p^n - n_e^n -\frac{\tau_n}{2} \nabla \cdot(D_e\nabla n_e^n ),
\label{semi2}
\end{align}
consequently,  $\phi^{n+1/2}$ and $\vec{E}^{n+1/2}$ can be reused in \eqref{e2} without additional calculation. Therefore, the elliptic equation is solved only once at each time step.

The basic idea for reducing the computational cost is to mimic the underlying mechanism of the second-order implicit midpoint rule \cite[Chapter 3]{iserles_2008}, in which the right-hand side appears only once at each time step.  To avoid solving nonlinear systems, this mechanism is applied only to the electric field; the other parts are implemented following the explicit midpoint method. Comparing the first-order scheme \eqref{semi-implicit} and our second-order scheme \eqref{e1}--\eqref{e2}, and focusing on the treatment of the electric field, the difference is similar to the difference between the backward Euler method and the implicit midpoint method. However, it is well known that the backward Euler method is L-stable, while the implicit midpoint method is not. Hence, due to the strong relation between L-stability and the asymptotic preserving property \cite{JinShi2009}, when using \eqref{e1}--\eqref{e2}, we will probably lose the asymptotic preserving property while gaining one additional numerical order. Nevertheless, due to its implicit nature, the scheme in \eqref{e1}--\eqref{e2} is indeed more stable than the explicit ones, as will be shown numerically in Section \ref{1dtest}.

It is worth noting that both (\ref{e15}) and (\ref{semi2}) are variable coefficient elliptic problems in which the coefficients vary at every time step during the streamer simulations. Thus, the coefficient matrix must be computed and assembled in each time step, whereas it needs to be calculated only once in the constant case. When a preconditioned iterative elliptic solver is used, the preconditioner must also be renewed in each step to solve the variable coefficient elliptic equation (again, this needs to be done only once in the constant case). The situation is similar if a direct solver is used. Therefore, in streamer simulations, solving a variable coefficient elliptic equation generally consumes more time than solving a Poisson equation with constant coefficients.

However, it is still not true to conclude that the second-order explicit scheme in \eqref{heun_ex_1}--\eqref{heun_ex_3} is faster than the second-order semi-implicit scheme in \eqref{e1}--\eqref{e2}. As we will show in Section \ref{differentmethods}, the semi-implicit scheme achieves better performance than explicit schemes in many Krylov elliptic solvers even under the same time steps. Moreover, the semi-implicit schemes remove the dielectric relaxation time restriction, which may allow a larger time step on many occasions to further shorten the total computational time. 

\subsection{Spatial discretization by FVM} \label{spatial}
Finite volume method is applied for spatial discretization. The computational domain is decomposed by a uniform grid with $M_x$, $M_y$, $M_z$ partitions in the $x$, $y$, $z$ directions, respectively. Therefore, the grid size is characterized by $\Delta x=(x_1-x_0)/M_x$, $\Delta y=(y_1-y_0)/M_y$, $\Delta z=(z_1-z_0)/M_z$. The grid cells are denoted by $I_{i,j,k}=[x_0+i \Delta x, x_0+(i+1)\Delta x] \times [y_0+j \Delta y, y_0+(j+1)\Delta y] \times [z_0+k \Delta z, z_0+(k+1)\Delta z]$, where $0 \leq i \leq M_x-1$, $0 \leq j \leq M_y-1$, and $0 \leq k \leq M_z-1$. The finite volume method is used for the spatial discretization, and we define
\begin{align}
(n_e)_{i,j,k}^n= \frac{1}{|I_{i,j,k}|} \int_{I_{i,j,k}} n_e^n(x,y,z)\mathrm{d}x \mathrm{d}y \mathrm{d}z.
\end{align}
Other notations, such as $(n_p)^n_{i,j,k}$ and $\phi^{n+1/2}_{i,j,k}$ are similarly defined. We adopt the classical second-order central scheme for \eqref{semi2}. Let $P_{i,j,k}^n$ be the discrete coefficient of the elliptic problem \eqref{semi2} defined by
\begin{align}
P_{i,j,k}^n=\frac{\varepsilon_0}{e} + \frac{\tau_n}{2} \left( \mu_p (n_p)^n_{i,j,k} + \mu_e (n_e)_{i,j,k}^n \right),
\end{align} 
and denote
\begin{displaymath}
P^n_{i\pm 1/2,j,k}=\frac{1}{2}(P^n_{i\pm 1,j,k}+P^n_{i,j,k}), \quad
P^n_{i,j\pm 1/2,k}=\frac{1}{2}(P^n_{i,j\pm 1,k}+P^n_{i,j,k}), \quad
P^n_{i,j,k\pm 1/2}=\frac{1}{2}(P^n_{i,j,k\pm 1}+P^n_{i,j,k}).
\end{displaymath}
Then, \eqref{semi2} is discretized as follows:
\begin{equation}
\begin{split}
&- \frac{P_{i+1/2,j,k}^n \Delta_{+x}\phi^{n+1/2} - P_{i-1/2,j,k}^n \Delta_{-x}\phi^{n+1/2}}{(\Delta x)^2} 
- \frac{P_{i,j+1/2,k}^n \Delta_{+y}\phi^{n+1/2} - P_{i,j-1/2,k}^n \Delta_{-y}\phi^{n+1/2}}{(\Delta y)^2} \\
&- \frac{P_{i,j,k+1/2}^n \Delta_{+z}\phi^{n+1/2} - P_{i,j,k-1/2}^n \Delta_{-z}\phi^{n+1/2}}{(\Delta z)^2} 
=  n_p^n - n_e^n - \frac{\Delta t^n}{2} \left( \frac{D_{e,x}}{(\Delta x)^2} \delta_x^2 n_e^n + \frac{D_{e,y}}{(\Delta y)^2} \delta_y^2 n_e^n + \frac{D_{e,z}}{(\Delta z)^2} \delta_z^2 n_e^n\right)
\end{split}
\label{e18}
\end{equation}
where the subscripts are neglected for the numerical solutions at $I_{i,j,k}$, e.g., $n_p^n$ denotes $(n_p)^n_{i,j,k}$, $\Delta_{+x} \phi^{n+1/2}$ and $\Delta_{-x} \phi^{n+1/2}$ denote the forward and backward differences of $\phi_{i,j,k}^{n+1/2}$ in the $x$ direction, respectively:
\begin{align}
\Delta_{+x} \phi^{n+1/2} = \phi_{i+1,j,k}^{n+1/2} - \phi_{i,j,k}^{n+1/2},\ \Delta_{-x} \phi^{n+1/2} = \phi_{i,j,k}^{n+1/2} - \phi_{i-1,j,k}^{n+1/2},
\end{align}
and similar notations are applied for $\Delta_{\pm y} \phi^{n+1/2}$ and $\Delta_{\pm z} \phi^{n+1/2}$; $\delta_x^2 n_e^n$ denotes the second-order central difference of $(n_e)^n_{i,j,k}$ in the $x$ direction:
\begin{align}
\delta_x^2 n_e^n = (n_e)^n_{i+1,j,k}-2(n_e)^n_{i,j,k} + (n_e)^n_{i-1,j,k},
\end{align}
and similar notations are used for $\delta_y^2 n_e^n$ and $\delta_z^2 n_e^n$. 

Afterwards, $\vec{E}$ can be calculated numerically by the central difference from the numerical solution of $\phi$, and $|\vec{E}|$ can be evaluated accordingly. In some cases \cite{sixcode,0022-3727-51-9-095206} where the mobility and diffusion coefficients depend on $| \vec{E} |$, interpolations can be applied to obtain $|\vec{E}|$ on the cell surfaces.

For the transport equations in \eqref{e1}--\eqref{e2}, the second-order MUSCL scheme combined with the Koren limiter is applied \cite{koren, van1979towards}. Ghost cells are used for all the boundary conditions of $n_e$ and $n_p$. This part of the spatial discretization is classical, and we omit the details here. Generally, we expect second-order accuracy from this spatial discretization for smooth solutions.

Due to the explicit treatment in the temporal discretization of the drift and diffusion terms, all temporal schemes in Section \ref{2nd_order} with the above FVM discretization should satisfy the following stability condition
\begin{align}
\tau_n \sum_{\alpha=x,y,z} \left( \frac{C_{\lambda} \mu_e |(E_\alpha)^{*}|_{\max}}{\Delta \alpha} + \frac{2D_{e,\alpha}}{(\Delta \alpha)^2} \right) \leq 1,
\label{timedrift}
\end{align}
where $(E_\alpha)^*$ denotes $E_{\alpha}^{n+1/2}$ for scheme \eqref{e1}--\eqref{e2}, $E_{\alpha}^{n}$ for explicit schemes, and $E_{\alpha}^{n+1}$ for scheme \eqref{semi-implicit}; $E_x$, $E_y$ and $E_z$ are the components of the electric field $\vec{E}=(E_x,E_y,E_z)^T$, and the subscript ``$\max$'' denotes the maximum value among all cells. 

We have two remarks on \eqref{timedrift}. Firstly, the problem is convection dominated, and typically the restriction on the time step determined by the convection term is stricter than that of the diffusion term. 
Secondly, the drift velocity of electrons $\mu_e |\vec E|$ is typically one or two orders of magnitude larger than that of positive ions $\mu_p |\vec E|$, and therefore only the stability condition for $n_e$ needs to be considered. 

The constant $C_{\lambda}$ in \eqref{timedrift} depends on both time integration and space discretization methods \cite{cockburn1989tvb,cockburn1998runge,shu1988total}. The stability of the schemes can be analyzed by fixing the electric field and neglecting the source terms. With the MUSCL finite volume method as the space discretization, von Neumann analysis (e.g. \cite[Chapter 20]{leveque2002finite}) indicates the second-order methods in this paper (including the proposed semi-implicit method and the second-order explicit scheme) are linearly stable without limiters under the condition \eqref{timedrift} with $C_{\lambda} = 1$. On the other hand, by Harten's theorem \cite{harten1983high}, it can be shown that the first-order methods \eqref{explicit} and \eqref{semi-implicit} are stable under \eqref{timedrift} with $C_{\lambda} =2$ provided that a proper slope limiter is applied, e.g., the Koren limiter which is used in this paper.

As a summary, we have discussed two constraints for the time step, one from the dielectric relaxation time \eqref{drt}, and the other from the convection and diffusion \eqref{timedrift}. For explicit schemes \eqref{explicit} and \eqref{heun_ex_1}--\eqref{heun_ex_3}, both conditions have to be satisfied; while for the proposed second-order semi-implicit \eqref{e1}--\eqref{e2} and the first-order semi-implicit scheme \eqref{semi-implicit}, it will be shown numerically in Section \ref{1dtest} that the constraint from the dielectric relaxation time can be relaxed. Which constraint is more restrictive depends on the problem setting and the spatial discretization. For the problems in which \eqref{timedrift} gives a stricter constraint, all the aforementioned schemes require similar time steps. However, for those problems where the dielectric relaxation constraint is tighter, semi-implicit schemes allow larger time steps, so that we can better match the errors of temporal and spatial discretization to maximize the computational efficiency.

\section{Multigrid preconditioned FGMRES elliptic solver}\label{solver}

To solve \eqref{e18}, an iterative solver is preferable to a direct solver. Although some state-of-the-art direct solvers retain the matrix sparsity to some degree, in three-dimensional simulations, a parallel direct solver still generally requires enormous amounts of memory, which is unaffordable when the number of DOFs becomes large and is therefore inapplicable. One example of using the parallel direct solver MUMPS can be found in \cite{0022-3727-51-9-095206}. 

In \cite{kacem2012full}, the geometric multigrid method was shown to be faster than the SOR method for solving the Poisson equation in 2D streamer discharges. Moreover, the convergence rate of the SOR method depends on the relaxation factor; however, maintaining its optimality at each time step is difficult because the coefficients in elliptic problem \eqref{e18} vary. 

We use a geometric multigrid as a preconditioner rather than a solver because the geometric multigrid preconditioned Krylov subspace solver may be more stable and efficient than using a geometric multigrid alone. In \cite{doi:10.1002/nla.1979}, a multigrid method was shown to be divergent for high-order FEMs when used as a solver, but convergence was achieved when the multigrid was combined with the conjugate gradient method. By investigating the eigenvalues of the iteration matrix, it was found in \cite{oosterlee1996use} that that while isolated large eigenvalues limit the convergence of a multigrid solver, the eigenvectors belonging to these large eigenvalues can be captured in a Krylov subspace constructed by GMRES within a few iterations, which accelerates the convergence of a multigrid solver. It was also shown in \cite{doi:10.1002/nla.1979, tatebe1993multigrid} that the multigrid preconditioner combined with the conjugate gradient method is faster and more stable than the multigrid solver alone.

\subsection{Preconditioned FGMRES solver}
Using a geometric multigrid as the preconditioner, we find that the geometric multigrid-preconditioned flexible generalized minimal residual (FGMRES) solver is the best among various Krylov subspace solvers, as discussed in Section \ref{differentmethods}. The flowchart of the preconditioned FGMRES is shown in Algorithm \ref{FGMRESalg} \cite{saad1993flexible}. Hereafter, the notation $\|\cdot \|$ denotes 2-norm.

\begin{algorithm}
	\caption{FGMRES with preconditioning to solve $A x = b$.}
	\label{FGMRESalg}
	\algsetup{indent=2em}
	\begin{algorithmic}[1]
        \STATE \textbf{1}. Initial guess $x_0$. Define a $(m+1) \times m$ zero matrix $\bar{H}_m=(h_{i,j})$, where $m$ is a given number indicating the dimension of Krylov subspace and $h_{i,j}$ denotes the $(i,j)$-th entry of $\bar{H}_m$. Let $k\leftarrow m$. 
		\STATE \textbf{2}. Arnoldi process:
		\STATE Compute $r_0 \leftarrow b - A x_0$, $\beta \leftarrow \| r_0 \|$ and $v_1 \leftarrow r_0 / \beta$.
		\FOR{$j = 1,\cdots,m$}
		\STATE Preconditioning: $z_j \leftarrow \bar{M}_j v_j$;
		\STATE Compute $\omega \leftarrow A z_j$;
		\FOR{$i=1,\cdots,j$}
		\STATE Gram-Schmidt process: $h_{i,j} \leftarrow (\omega, v_i)$, $\omega \leftarrow \omega - h_{i,j}v_i$;
		\ENDFOR
		\STATE Compute $h_{j+1,j} \leftarrow \| \omega \|$, $v_{j+1} \leftarrow \omega / h_{j+1,j}$;
		\STATE Compute the residual $r_j \leftarrow \min_{y} \| \beta e_j - \bar{H}_j y \|$, where $\bar{H}_j$ is the upper left $(j+1) \times j$ sub-matrix of $\bar{H}_m$ and $e_j = [1,0,\cdots,0]^{T}$ with totally $j+1$ entries;
		\STATE Check the stopping criterion. If satisfied, let $k \leftarrow j$ and go to line 14.
		\ENDFOR
		\STATE Define a matrix $Z_k \leftarrow [z_1,\cdots,z_k]$.
		\STATE \textbf{3}. Form the iterative solution $x_k \leftarrow x_0 + Z_k y_k$, where $y_k = \arg \min _y \| \beta e_k - \bar{H}_k y \|$.
		\STATE \textbf{4}. Restart: If the stopping criterion is not satisfied, let $x_0 \leftarrow x_k$ and $k \leftarrow m$; go to line 2.
	\end{algorithmic}
\end{algorithm}

As shown in line 5 of Algorithm \ref{FGMRESalg}, for different basis vectors $v_j$, different preconditioning matrices $\bar{M}_j$ can be selected, which provides the “flexibility” reflected in the solver's name. The price is that the preconditioned vectors $z_j$ in line 5 must be stored to form the matrix $Z_m$, resulting in a larger memory cost than is achieved by the classical generalized minimal residual (GMRES) method which stores only the vectors $v_j$. However, the flexibility resulting from different preconditioners helps to improve the robustness of the GMRES algorithm, as shown in \cite{saad1993flexible}. In our FGMRES implementation, we initially set $m=30$ and selected the multigrid as the preconditioner in line 5.

\subsection{Multigrid preconditioner} \label{multigrid}
A geometric multigrid preconditioner is chosen to accelerate the convergence of the FGMRES solver. 

Our implementation of the geometric multigrid preconditioner uses a full multigrid (FMG) \cite{kacem2012full} for the first time step, and V-cycle multigrid afterwards. In the first time step, no previous information is available, and therefore we simply make a zero initial guess, with the expectation that FMG will achieve faster convergence. Subsequently, the potential $\phi$ calculated during the previous time step is adopted as the initial guess. Given a good initial guess, the cheaper V-cycle multigrid gives better performance.

\begin{figure}[htbp]
	\centering
	\includegraphics[width=\textwidth]{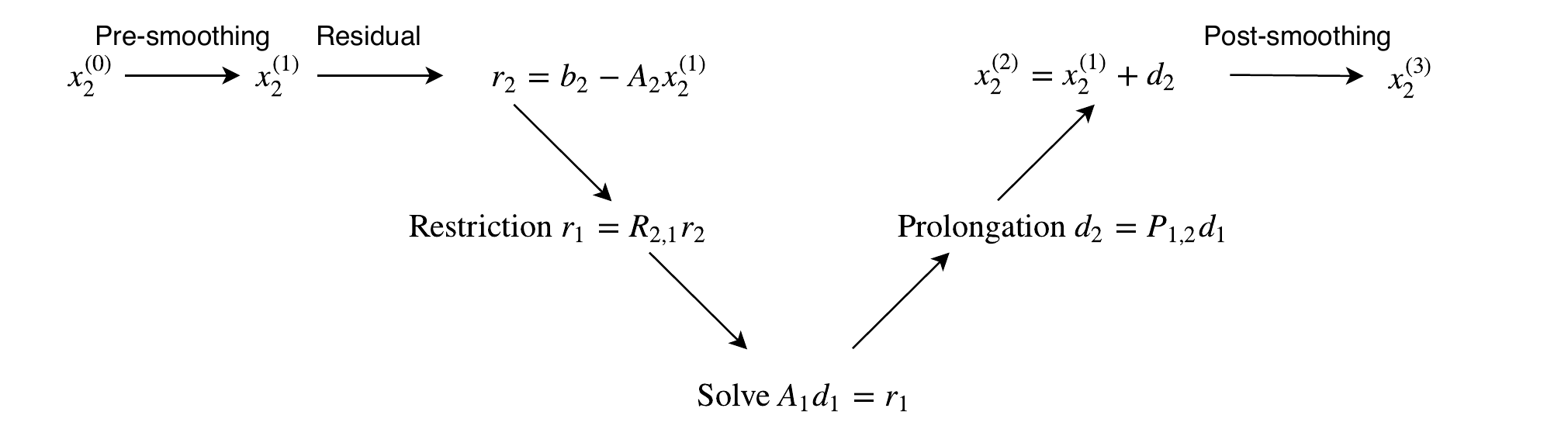}
	\includegraphics[width=0.7\textwidth]{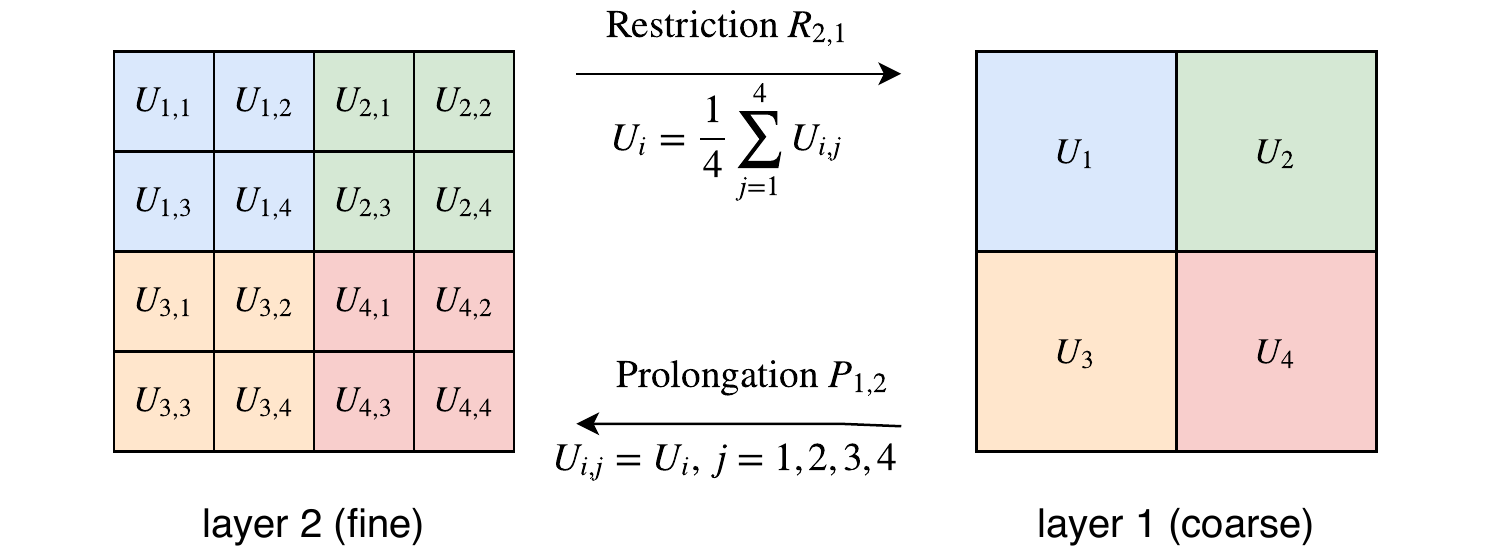}
	\caption{Diagram of V-cycle geometric multigrid for uniform mesh in two layers.}
	\label{multigriddiagram}
\end{figure}

To introduce the multigrid preconditioner, we first provide a simple review of the multigrid solver \cite[Chapter 2]{trottenberg2000multigrid}. In the following, assume that the elliptic equation on grid level $l$ is discretized as follows:
\begin{align}
A_l x_l = b_l.
\label{e31}
\end{align} 
A diagram showing the procedure of the two-layer V-cycle multigrid method is given in Figure \ref{multigriddiagram}, where the subscripts $2$ and $1$ denote the second layer (the fine layer) and the first layer (the coarse layer) respectively. The restriction and prolongation are shown using a 2D example of $4\times 4$ and $2 \times 2$ meshes. When solving the equation $A_1 d_1 = r_1$  shown at the bottom of the V-cycle, this multigrid procedure can be called recursively, resulting in a multi-layer multigrid solver.

Some smoothers commonly used in sequential computation include the Gauss-Seidel method and the successive over relaxation (SOR) method \cite{tatebe1993multigrid}. However, when parallelized, the efficiency of these methods is impaired due to their sequential nature. Therefore, we adopt the Chebyshev smoother in our implementation, which is a polynomial smoother based on Chebyshev polynomials. The performance of polynomial smoothers (including Chebyshev polynomials) and the parallel Gauss-Seidel smoother were compared in \cite{ADAMS2003593}; the results show that polynomial smoothers are preferable in a parallel environment. In general, given a polynomial $p_n(x)$ of degree $n$, the associated polynomial smoother in the pre-smoothing of the multigrid algorithm is
\begin{align}
x_2^{(1)} & = x_2^{(0)} + p_n(A_2)(b_2-A_2x_2^{(0)}),
\label{smootheq}
\end{align}
which smooths out the error as follows:
\begin{displaymath}
x_2^{(1)} - A_2^{-1} b_2 = q_{n+1}(A_2) (x_2^{(0)} - A_2^{-1} b_2),
\end{displaymath}
where $q_{n+1}(x) = 1 - x p_n(x)$. The Chebyshev smoother uses the following polynomials \cite{brannick2015}:
\begin{align}
q_{n+1}(x) = T_{n+1}\left( \frac{\lambda_{\max}(A_2)+ \lambda^*(A_2) - 2x}{\lambda_{\max}(A_2)-\lambda^*(A_2)} \right) \bigg/ T_{n+1}\left( \frac{\lambda_{\max}(A_2)+ \lambda^*(A_2)}{\lambda_{\max}(A_2)-\lambda^*(A_2)} \right),
\label{qntn}
\end{align}
where $\lambda_{\max}(A_2)$ is the largest eigenvalue of $A_2$ and is usually replaced by an approximation of the largest eigenvalue in practice; $\lambda^*(A_2)$ is selected manually, both of which will be discussed later in this section; and $T_{n+1}(x)$ is the Chebyshev polynomial of degree $n+1$. $p_n(x)$ can be obtained accordingly. By introducing the two matrices $P_2=p_n(A_2)$ and $Q_2 = q_{n+1}(A_2)$, we can re-write the pre-smoothing operation in \eqref{smootheq} as follows:
\begin{align}
x_2^{(1)} = Q_2 x_2^{(0)} + P_2 b_2.
\label{e19}
\end{align}
Moreover, the Chebyshev iteration can be further improved to become a “preconditioned Chebyshev iteration” by introducing another preconditioner on top of it, for which we refer the readers to \cite{brannick2015} for details.

The post-smoothing operation in Figure \ref{multigriddiagram} is applied to $x_2^{(2)}$ in the same way as \eqref{e19}. Therefore, the V-cycle multigrid in Figure \ref{multigriddiagram} is formulated as follows:
\begin{align}
x_2^{(3)} = x_2^{(0)} + M_2 (b_2 - A_2 x_2^{(0)}),
\label{eRichard}
\end{align}
where $M_2=Q_2P_2 + P_2 + Q_2 P_{1,2}A_1^{-1}R_{2,1}(I_2-A_2P_2)$ and $I_2$ is the identical matrix on the second layer. The derivation of \eqref{eRichard} and the corresponding equation for the general multi-layer V-cycle method are shown in Appendix \ref{appendixa}. 

In our implementation, the quadratic-polynomial preconditioned Chebyshev smoother is applied to both pre-smoothing and post-smoothing, using the one-step local symmetric successive over relaxation method (SSOR) as the preconditioner. The values of $\lambda_{\max}(A_l)$ and $\lambda^*(A_l)$, which are required in the Chebyshev iteration (see \eqref{qntn}), are estimated by
\begin{displaymath}
\lambda_{\max}(A_l) \approx 1.1 \lambda_{\max}(H_m), \qquad
\lambda^*(A_l) = 0.1 \lambda_{\max}(H_m),
\end{displaymath}
where $H_m$ is the upper Hessenberg matrix $(h_{i,j})_{m\times m}$ obtained by applying the Arnoldi process (without preconditioning) to $A_l$, as shown in Algorithm \ref{FGMRESalg}. We use $m = 10$  for the eigenvalue estimate. A sequential direct solver is applied to the coarsest mesh. 

When the multigrid is used in the ``Preconditioning'' step in line 5 of Algorithm \ref{FGMRESalg}, we apply the multigrid preconditioner $\bar{M}_j$ to a vector $v_j$ by setting a zero initial $x_2^{(0)}=\vec{0}$ and $b_2 = v_j$ in the multigrid solver, and run the V-cycle once. The output will be the preconditioned vector $z_j = \bar{M}_j v_j$, turning a multigrid solver into a preconditioner for Krylov subspace solvers. More details of the preconditioning process can be found in, e.g., \cite{ADAMS2003593, tatebe1993multigrid}. 


\section{MPI based parallel implementation}\label{mpiparallel} 
Most High Performance Computing (HPC) platforms support MPI, and a variety of implementations are available. MPI is responsible for the communication between different processes by sending and receiving messages, and supports parallel computing using thousands of cores, to fully utilize the power of modern clusters.

In our implementation, we partition the 3D grid into Cartesian subgrids of equal size in each direction. Each process stores only the portion of the solutions defined in one of the subgrids. To reduce the communication latency, for each subgrid, the number of cells in each direction is approximately equal. 

To apply the MUSCL scheme, each process requires inter-process communication to retrieve the values of $n_e$ and $n_p$ on the two adjacent layers of cells from its neighbouring processes. For a 3D uniform grid, an interior subgrid whose size is $M \times N \times P$ should receive $4(MN + NP + MP)$ cells of unknowns from the surrounding processes and should send the same number of unknowns to them. Therefore, the local communication/computation ratio can be characterized as following:
\begin{align}
\frac{\text{communication}}{\text{computation}} = \frac{4(MN + NP + MP)}{MNP} \leq \frac{12}{\min\{ M,N,P\}},
\end{align}
showing that the communication cost is one order of magnitude lower than the computation for interior processes. We use ghost cells to address the boundary conditions; therefore, the communication required for processes handling boundary conditions is less than the communication required for interior processes.

The communication for the potential $\phi$ is similar, but the communication stripes have a width of only one cell in most layers of the multigrid. The only exception is that at the coarsest layer of the multigrid preconditioner where a direct solver is used, gathering and broadcasting operations are still needed for a small amount of data.

\section{Accuracy and stability as well as efficiency test}\label{TestSection}
In this section, we first adopt a 1D dimensionless model to illustrate the convergence order and stability of our second-order semi-implicit scheme. Then, we use a 3D problem to show the scalability of our simulator, and make a comparison between several algebraic elliptic solvers.

\subsection{Comparison on convergence and stability}\label{1dtest}
The following dimensionless model problem in 1D, which has a form similar to \eqref{eM}, is adopted to show that the proposed semi-implicit scheme is second-order accurate in time, and is more stable than the explicit schemes:
\begin{align}
\left\{ \begin{aligned}
& \partial_t n_e - \partial_x (\mu_e E n_e) - D_e \partial_{xx}(n_e) =  S \exp({-K/|E|}) \mu_e |E| n_e, \\
& \partial_t n_p + \partial_x (\mu_p E n_p) =  S \exp({-K/|E|}) \mu_e |E| n_e, \qquad \qquad \qquad x \in I, \ \  t > 0,\\
& -\gamma \partial_{xx}\phi = n_p - n_e, \ E = - \partial_x \phi,
\end{aligned}
\right.
\label{eT}
\end{align}
where $I = (0,1)$. Dirichlet boundary conditions $\phi(0,t)=1$ and $\phi(1,t)=0$ are applied for $\phi=\phi(x,t)$, while homogeneous Neumann boundary conditions are applied at all boundaries for $n_e=n_e(x,t)$ and at inflow boundary for $n_p=n_p(x,t)$. Parameters in \eqref{eT} are given by $\mu_e=1$, $\mu_p=0.09$, $D_e = 10^{-4}$, $S=1000$ and $K=4$, while constant $\gamma$ will be given later.  Constant time steps $\tau_n = \tau$ are used, and the computation is performed until $T=0.05$. It should be mentioned that the dimensionless dielectric relaxation time constraint for \eqref{eT} is
\begin{equation}
\tau \leq \tau_{\text{diel}} = \frac{\gamma}{\max(\mu_p n_p + \mu_e n_e)}.
\label{1ddiel}
\end{equation}
Different temporal schemes introduced in Section \ref{2nd_order} are implemented with the same spatial discretization.

\paragraph{Study of convergence}\label{convergence}
In this testing example, we set $\gamma = 10^{-3}$ in \eqref{eT}. The initial value is $n_e(x,t=0) = n_p(x,t=0) = 10^{-6} + 0.1\exp(- 100(x-0.5)^2)$. For all the calculations in this example, we fix the ratio of the time step to the grid size at $\tau / \Delta x = 0.25$. The finite volume method with unlimited linear reconstruction is applied for spatial discretization. The ``exact solution" for $(n_e)_{\text{ref}}$ and $(n_p)_{\text{ref}}$ are calculated by second-order explicit scheme \eqref{heun_ex_1}--\eqref{heun_ex_3} with $\tau = 0.005/2^{8}$ which is sufficiently small. The numeric results are given in Tables \ref{t5} and \ref{t6}.
\begin{table}[htbp]
\caption{$L^2$-norm error of the second-order semi-implicit scheme (\ref{e1})--(\ref{e2}) in the 1D testing problem.}
\centering
\begin{tabular}{l l l l l l l}
\hline
$\Delta t$ & $0.005$ & $0.005/2$ & $0.005/2^2$ & $0.005/2^3$ & $0.005/2^4$ & $0.005/2^5$ \\
\hline
$\| n_e - (n_e)_{\text{ref}}\|$ & 3.1660$\times 10^{-4}$ & 8.3832$\times 10^{-5}$ & 2.1650$\times 10^{-5}$ & 5.5097$\times 10^{-6}$ & 1.3876$\times 10^{-6}$ & 3.4508$\times 10^{-7}$ \\
order & -- & 1.9171 & 1.9531 & 1.9743 & 1.9894 & 2.0075 \\
\hline
$\| n_p - (n_p)_{\text{ref}}\|$ & 2.5303$\times 10^{-4}$ & 6.3421$\times 10^{-5}$ & 1.6028$\times 10^{-5}$ & 4.0412$\times 10^{-6}$ & 1.0134$\times 10^{-6}$ & 2.5158$\times 10^{-7}$ \\
order & -- & 1.9962 & 1.9844 & 1.9877 & 1.9956 & 2.0101 \\
\hline
\end{tabular}
\label{t5}
\end{table}

\begin{table}[htbp]
\caption{$L^2$-norm error of the first-order semi-implicit scheme \eqref{semi-implicit} in the 1D testing problem. }
\centering
\begin{tabular}{l l l l l l l}
\hline
$\Delta t$ & $0.005$ & $0.005/2$ & $0.005/2^2$ & $0.005/2^3$ & $0.005/2^4$ & $0.005/2^5$ \\
\hline
$\| n_e - (n_e)_{\text{ref}}\|$ & 1.7405$\times 10^{-3}$ & 9.9821$\times 10^{-4}$ & 5.3935$\times 10^{-4}$ & 2.8098$\times 10^{-4}$ & 1.4349$\times 10^{-4}$ & 7.2514$\times 10^{-5}$ \\
order & -- & 0.8021 & 0.8881 & 0.9408 & 0.9696 & 0.9846 \\
\hline
$\| n_p - (n_p)_{\text{ref}}\|$ & 1.5186$\times 10^{-3}$ & 8.8187$\times 10^{-4}$ & 4.8021$\times 10^{-4}$ & 2.5122$\times 10^{-4}$ & 1.2856$\times 10^{-4}$ & 6.5044$\times 10^{-5}$ \\
order & -- & 0.7841 & 0.8769 & 0.9347 & 0.9665 & 0.9830 \\
\hline
\end{tabular}
\label{t6}
\end{table}

The results in Tables \ref{t5} and \ref{t6} clearly demonstrate that the proposed semi-implicit scheme \eqref{e1}--\eqref{e2} is indeed second order time accurate, while the previously used semi-implicit scheme \eqref{semi-implicit} is only first order time accurate, though second-order spatial discretization is used. 

It is worth emphasizing that the proposed second-order semi-implicit scheme needs to solve the elliptic equation only once during each time step, the same as the first-order semi-implicit scheme. To gain second order time accuracy, the only additional cost is an explicit stage for $n_p$ and $n_e$ at each time step, which is relatively cheap compared with solving the elliptic equation.

\paragraph{Study of stability in terms of the dielectric relaxation time restriction}
To show that the semi-implicit scheme \eqref{e1}--\eqref{e2} is able to alleviate the dielectric relaxation time restriction, and is thus more stable than explicit schemes, we perform the calculation with $\gamma = 10^{-5}$ in \eqref{1dtest}. The initial value is $n_e(x,t=0) = n_p(x,t=0) = 10^{-6} + \exp(- 100  (x-0.5)^2)$. Koren limiter is applied in finite volume discretization.

Note the goal of this example is to check the instability raised by the dielectric relaxation time restriction. As discussed at the end of Section \ref{NumericalMethod}, the constraints of the time step include the dielectric relaxation time constraint and the CFL-type stability condition. In this 1D setting, they can be represented, respectively, by \eqref{1ddiel} and 
\begin{align}
\tau \leq \tau_{\text{CFL}} = \frac{\Delta x}{C_\lambda \mu_e |E|_{\max} + 2 D_e / \Delta x}.
\label{1ddrift}
\end{align}
To get a good estimation of $\tau_{\text{diel}}$ and $\tau_{\text{CFL}}$ for this test problem, we first perform the simulation on a very fine mesh $\Delta x = 1 / 12800$ with $\tau = 1 / 128000$, using second-order explicit scheme \eqref{heun_ex_1}--\eqref{heun_ex_3}, and record the maximum values of $|E|$ and $(\mu_p n_p + \mu_e n_e)$ throughout the simulation. Such results are considered to have sufficient accuracy, so that we can use these values to get a precise estimation of $\tau_{\text{diel}}$ defined in \eqref{1ddiel}, and the result is $\tau_{\text{diel}} = 9.1736 \times 10^{-6}$. To estimate $\tau_{\text{CFL}}$, we insert the estimated value of $|E|_{\max}$ into \eqref{1ddrift}, with $C_{\lambda}$ set to be $2$ and $\Delta x$ chosen as a relatively larger cell size $\Delta x = 1/400$, which yields $\tau_{\text{CFL}} = 4.7448 \times 10^{-4}$. The numerical tests presented below will be carried out on the uniform grid with $\Delta x = 1/400$. Thus we have $\tau_{\text{CFL}} \approx 52 \tau_{\text{diel}}$, meaning that a much better efficiency can be achieved if we can break the dielectric relaxation time constraint.


Five different time steps, i.e., $0.5\tau_{\text{diel}}$, $\tau_{\text{diel}}$, $3\tau_{\text{diel}}$, $10\tau_{\text{diel}}$ and $50 \tau_{\text{diel}}$, are used to test the stability. All these five time steps are less than $\tau_{\text{CFL}}$, and stability condition \eqref{1ddrift} is always satisfied for all simulations before numerical blow-up occurs. We consider a simulation to be unstable if $n_e>10$ is detected in this example. According to our experiments, this condition always leads to a quick numerical blow-up of the solution. In addition to the proposed semi-implicit scheme \eqref{e1}--\eqref{e2}, we implemented three other temporal discretizations for comparisons: the first-order explicit scheme \eqref{explicit}, the first-order semi-implicit scheme \eqref{semi-implicit}, and second-order explicit scheme \eqref{heun_ex_1}--\eqref{heun_ex_3}.

\begin{table}[htbp]
\caption{Stability of different temporal discretizations on a 1D testing problem. }
\centering
\begin{tabular}{l|l|l|l|l|l}
	\hline
	Temporal scheme & $\tau = 0.5 \tau_{\text{diel}}$ & $\tau = \tau_{\text{diel}}$ & $\tau = 3\tau_{\text{diel}}$ & $\tau = 10\tau_{\text{diel}}$ & $\tau = 50 \tau_{\text{diel}}$ \\
	\hline
	$2^{\text{nd}}$ order semi-implicit \eqref{e1}--\eqref{e2} & stable & stable & stable & stable & stable \\
	$1^{\text{st}}$ order semi-implicit \eqref{semi-implicit} & stable & stable & stable & stable & stable \\
	$2^{\text{nd}}$ order explicit \eqref{heun_ex_1}--\eqref{heun_ex_3}  & stable & stable  & unstable & unstable & unstable \\
	$1^{\text{st}}$ order explicit \eqref{explicit} & stable & stable & unstable & unstable & unstable  \\
	\hline
\end{tabular}
\label{tstable}
\end{table}

The results in Table \ref{tstable} clearly show that the two semi-implicit methods remain stable when the time step exceeds $\tau_{\mathrm{diel}}$ and reaches $50 \tau_{\mathrm{diel}}$. In contrast, the two explicit methods exhibit instability. As indicated, the stability condition \eqref{1ddrift} is still fulfilled for all the time steps and schemes. Therefore the instability is caused by the violation of the dielectric relaxation time constraint, and the semi-implicit schemes truly allow larger time steps on this occasion. 

Although we focus on the stability in this example, we can calculate the $L^2$ errors for the stable schemes in Table \ref{tstable} using the numerical solution on the fine mesh as the reference. The errors range from 8.3245$\times 10^{-5}$ to 5.4554$\times 10^{-4}$ for the second-order semi-implicit scheme \eqref{e1}--\eqref{e2}, and from 1.8287$\times 10^{-4}$ to 9.0213$\times 10^{-3}$ for the first-order semi-implicit scheme \eqref{semi-implicit}, indicating that all stable results in Table \ref{tstable} still make reasonable predictions even when a relatively large time step $\tau = 50 \tau_{\text{diel}}$ is used with a coarse mesh size $\Delta x  = 1/400$.

As indicated in Section \ref{spatial}, the dielectric relaxation time constraint is not always the tightest time step constraint generally. However, the newly proposed semi-implicit method provides an alternative other than the explicit schemes. In addition, it requires solving the elliptic equation only once during each time step, while the explicit scheme \eqref{heun_ex_1}--\eqref{heun_ex_3} requires twice, which outweigh its possible drawback in the slower computation of the variable coefficient elliptic equation, as will be shown in Section \ref{differentmethods}.

\subsection{Scalability of MPI parallelization}\label{scalability}
The scalability means the ability to reduce the execution time as the number of cores increases. Scalability can be measured by speedup $S_p$, which is the ratio of the execution time of the sequential program to the execution time of the parallel program over $p$ processes. 

We developed our codes based on the well-known PETSc \cite{petsc-web-page}. PETSc contains data structures and routines for both scalable and parallel solutions of partial differential equations, and it supports MPI parallelism. The simulations were performed on the cluster Tianhe2-JK located at Beijing Computational Science Research Center. It includes 514 computational nodes, each of which is equipped with two Intel Xeon E5-2660 v3 CPUs (10 cores, 2.6 GHz) and 192 GB of memory. The nodes are connected by a TH high-speed network interface. More details can be found at {\tt https://www.csrc.ac.cn/en/facility/cmpt/2015-05-07/8.html}.
	
	Unless otherwise stated, we used the following setup for a double-headed streamer in a homogeneous field between two parallel planes at atmospheric pressure $P = 760$ Torr in following testings and applications, 
	\begin{align}
	\tilde{n}(\vec{x}) = 10^8 + 10^{14} \exp \left( - \left(\frac{z-0.5}{\sigma_z} \right)^2 - \left( \frac{(x-1)^2 + (y-1)^2}{\sigma_r^2} \right) \right)~ \mbox{cm}^{-3},
	\label{eq51}
	\end{align}
	where $\sigma_z=0.027$ and $\sigma_r = 0.021$. The voltage applied is $\phi|_{z=z_1}=\phi_0=52 \text{~kV}$. We adopt the same parameters as in \cite{movingmesh2007, dhali1987two} hereafter: $\mu_e=-2.9\times 10^5/P$ cm$^2/$(Vs) and $\mu_p=2.6 \times 10^3/P$ cm$^2/$(Vs), respectively;  $\alpha(|\vec{E}|) = 5.7P \exp(-260P/|\vec{E}|)$ cm$^{-1}$; and $D_e =\diag (D_{e,x},D_{e,y},D_{e,z})=\diag(2190,2190,1800)$ cm$^2/$s. 
	
	The convergence criterion for the iterative algebraic elliptic solver $A \tilde{\phi} = \beta$ is given by a tolerance of the relative residual, i.e., the iteration continues until
	\begin{align}
	\frac{\| A \tilde{\phi} - \beta \|_2}{\| \beta \|_2} < \varepsilon,
	\label{criterion}
	\end{align}
	where the tolerance $\varepsilon$ is set to $\varepsilon=10^{-8}$ in all the simulations hereafter.

The scalability of our simulator, which is the second-order semi-implicit scheme \eqref{e1}--\eqref{e2} combined with multigrid preconditioned FGMRES elliptic solver, is tested. We consider a domain $\Omega = (0,1)\times(0,1)\times(0,1)$ cm$^3$, with three different mesh sizes: $256\times256\times320$, $512\times512\times640$ and $1024\times1024\times1280$. The time step is chosen to be proportional to the mesh size, which is $\tau_n = \Delta z/v_{\mathrm{ch}}$ with $v_{\mathrm{ch}}$ being the maximum characteristic speed. Here we choose $v_{\mathrm{ch}}=3| \mu_e E_z |$ and $E_z = 208$ kV$\cdot$cm$^{-1}$ to ensure stability. Using a fixed time step $\tau=\tau_n$, we execute the code for 50 time steps, and record the elapsed wall-clock time for the whole run. Additionally, to obtain a more reliable result, we execute the same code five times and take the average at each mesh size.

The code is executed over different numbers of nodes, using all 20 cores in each node. This mode (in which all available cores are used in each node) is called the ``compact mode'' in \cite{0022-3727-51-9-095206}. The average elapsed times are given in Table \ref{t256}. 
 
\begin{figure}[htbp]
	\centering
        \begin{subfigure}[b]{0.32\textwidth}
		\includegraphics[width=\textwidth]{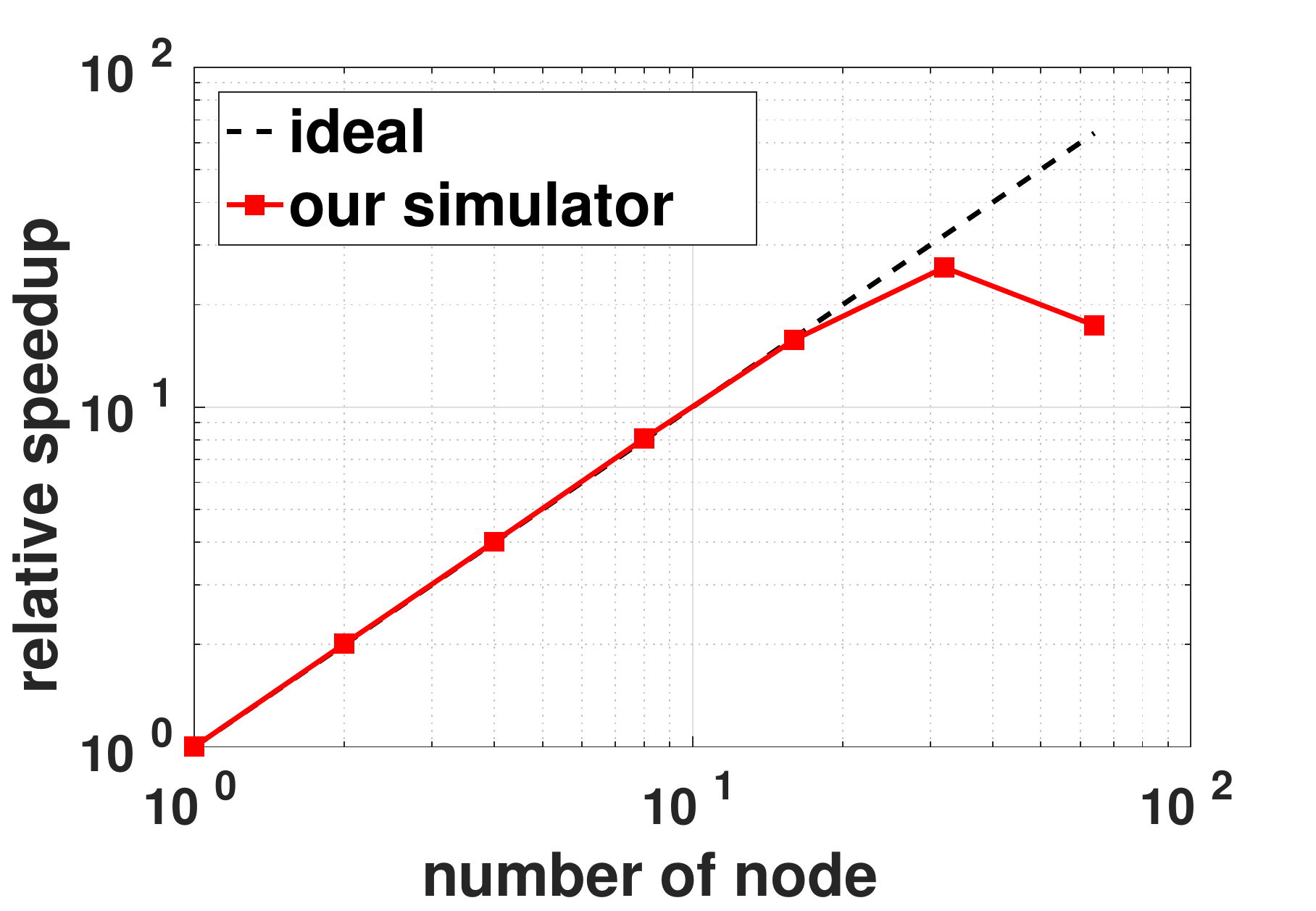}
		\caption{mesh size $256\times 256 \times 320$} \label{scala1}
	\end{subfigure}
	\begin{subfigure}[b]{0.32\textwidth}
		\includegraphics[width=\textwidth]{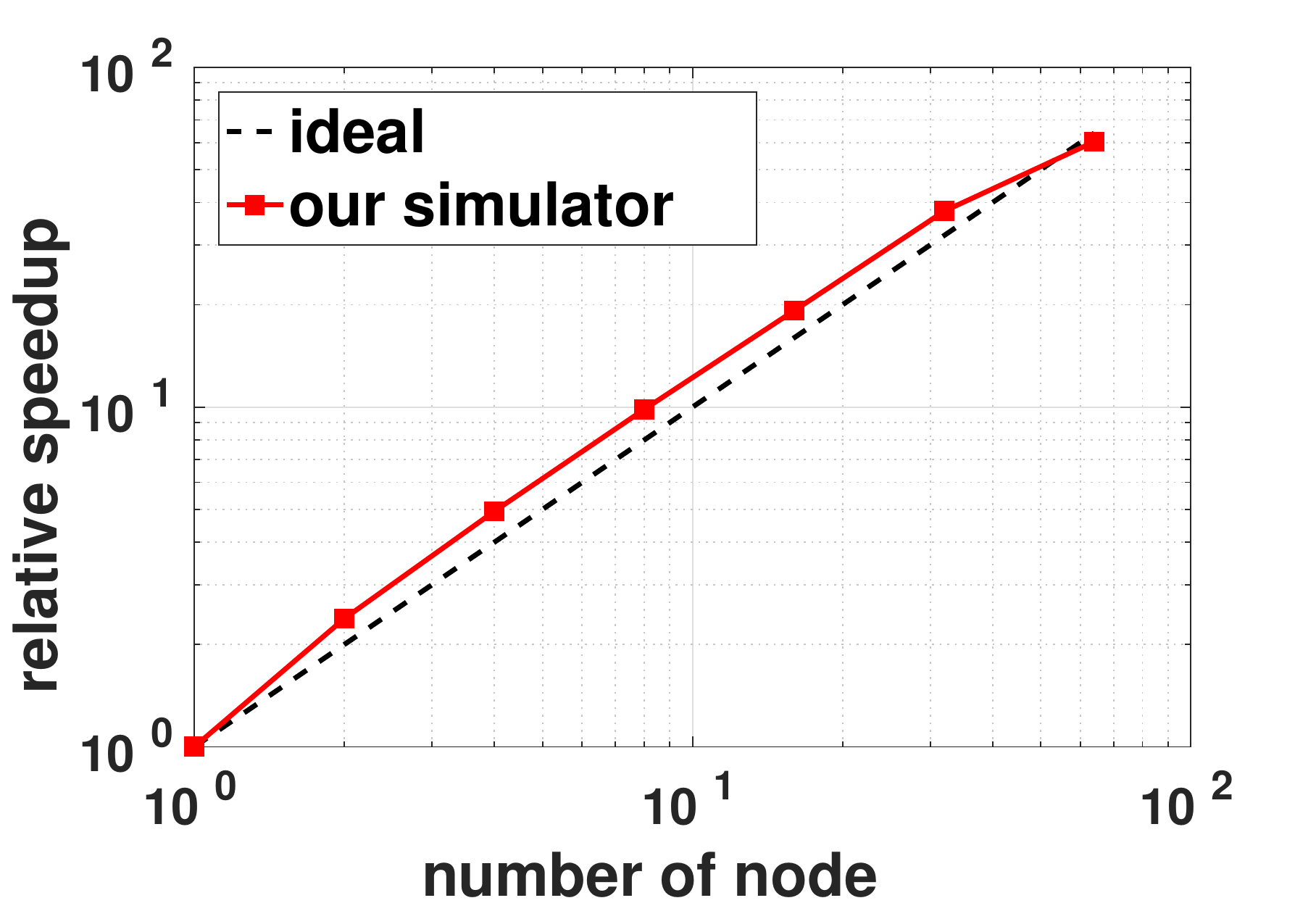}
		\caption{mesh size $512\times 512 \times 640$} \label{scala2}
	\end{subfigure}
	\begin{subfigure}[b]{0.32\textwidth}
		\includegraphics[width=\textwidth]{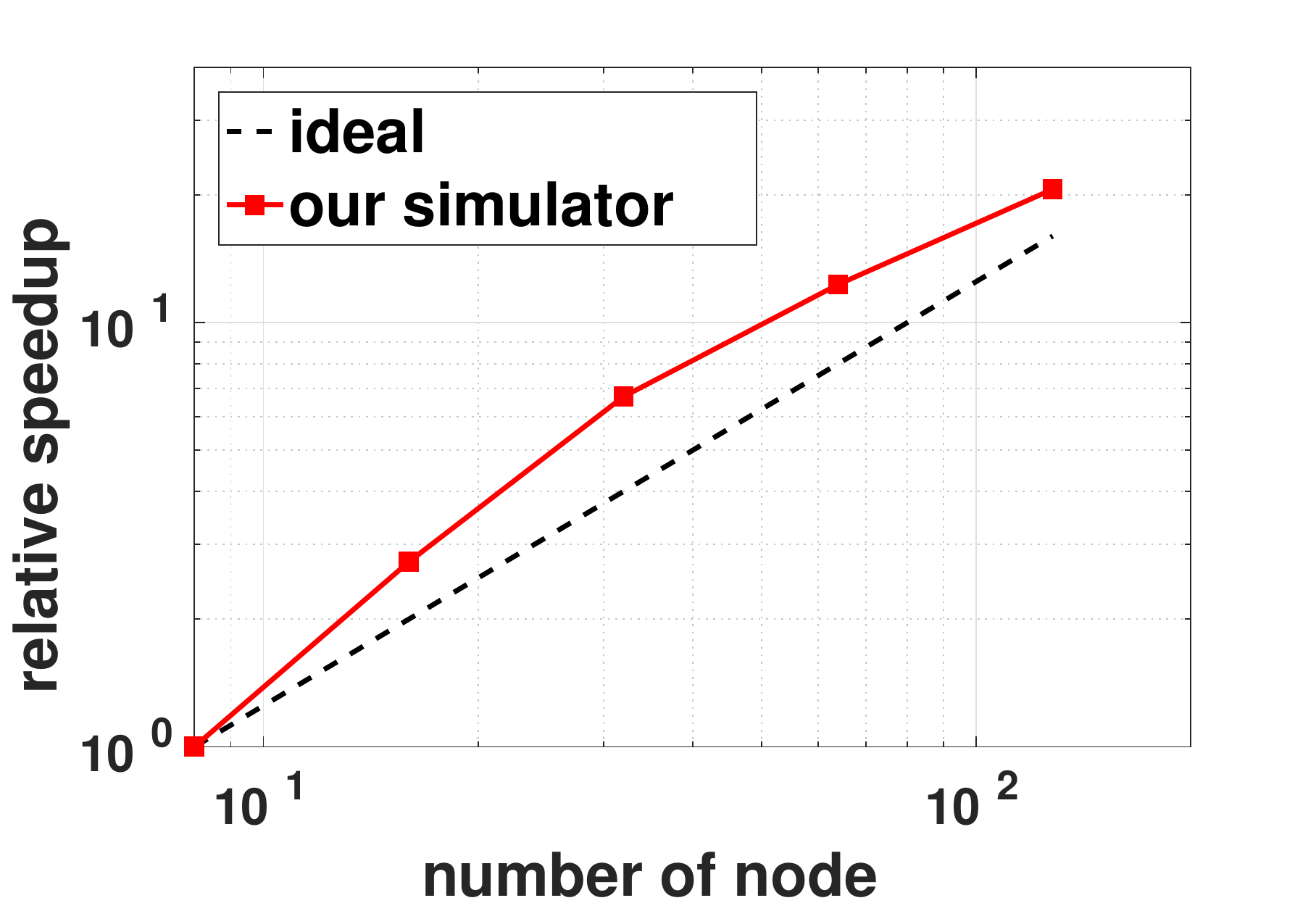}
		\caption{mesh size $1024\times 1024 \times 1280$} \label{scala3}
	\end{subfigure}
        \caption{Scalability of the second-order semi-implicit scheme using the multigrid preconditioned FGMRES solver over three meshes. Summarized from Table \ref{t256}.}
	\label{fscala}
\end{figure}

Note that the times shown in the tables are the average values over five runs, each with 50 time steps, instead of the average times for a single time step. The data are summarized in Figure \ref{fscala} for clarity, where relative speedup denotes the speedup with respect to the execution time using the smallest number of nodes in Table \ref{t256}.

\begin{table}[htbp]
	\caption{Mean time for 50 time steps on three different meshes, using {second-order semi-implicit scheme with multigrid preconditioned FGMRES, with} 20 cores in each node.}
	\centering
	\begin{tabular}{l l l l l l l l l}
		\hline
		Number of nodes & 1 & 2 & 4 & 8 & 16 & 32 & 64 & 128\\
		\hline
		Mesh size: $256\times256\times320$ & 299.94 & 149.31 & 74.743 & 37.093 & 19.003 & 11.645 & 17.234 & --\\
		\hline
		Mesh size: $512\times512\times640$ & 3007.2 & 1261.3 & 609.70 & 305.39 & 156.41 & 79.722 & 49.806 & --\\
		\hline
		Mesh size: $1024\times1024\times1280$ & -- & -- & -- & 3753.5 & 1375.1 & 560.89 & 305.13 & 181.97\\
		\hline
	\end{tabular}
	\label{t256}
\end{table}

Generally the satisfactory scalability of our program can be seen in Figure \ref{fscala}. Figure \ref{scala1} shows a nearly linear speedup when the number of nodes is small. When 32 or more nodes are used, the speedup obviously becomes sublinear. In particular, the time consumed for 64 nodes is even larger than that for 32 nodes. The reason is that the ratio of communication to computation becomes larger as the number of nodes increases. This tendency is visible in Figures \ref{scala2} and \ref{scala3} even though the larger computational work load postpones the significant drop of the parallel efficiency. It is interesting that in these two figures, we sometimes obtain a performance even better than the ideal case. One possible explanation for this phenomenon is that when the number of nodes is small, each node is heavily loaded, causing a lower cache hit ratio \cite[Overview \& Chapter 1]{cachebook}; while for each process, the amount of data required for communication is relatively large, causing more network latency. Such a phenomenon can also be observed clearly in Table \ref{t256}.

\subsection{Comparison of different algebraic elliptic solvers} \label{differentmethods}
In this subsection, we first study the performance of multigrid preconditioned FGMRES solver and then compare it with R\&B SOR and other multigrid preconditioned Krylov subspace methods.

We again adopt the double-headed streamer in homogeneous field for testing purposes, using the same configuration as in Section \ref{scalability} and three different mesh sizes $256\times256\times320$, $512\times512\times640$ and $1024\times1024\times1280$. As mentioned in Section \ref{multigrid}, a zero initial guess and the FMG preconditioner are used in the first time step; subsequently, the V-cycle multigrid preconditioner is applied. We simulate the double-headed streamer until $2.5$ ns, using a fixed time step of $\tau_n = 2$ ps. Therefore, $1250$ time steps are required to finish the simulation. 

We execute our program on 640 cores, distributed among 32 nodes, with 20 cores on each node. We record the maximum wall-clock time consumed by the elliptic solver over all cores, including both the computation and communication in the solver as well as the assembly of the coefficient matrix and the right-hand side. The total times consumed by the elliptic solver are 320.76 s, 1813.2 s and 14001 s for the three aforementioned mesh sizes, respectively, and these do not exceed linear growth with the number of DOFs. Note that these times do not include the computation for quantities other than $\phi$.

\begin{figure}[htbp]
	\centering
	\begin{subfigure}[b]{0.32\textwidth}
		\includegraphics[width=\textwidth]{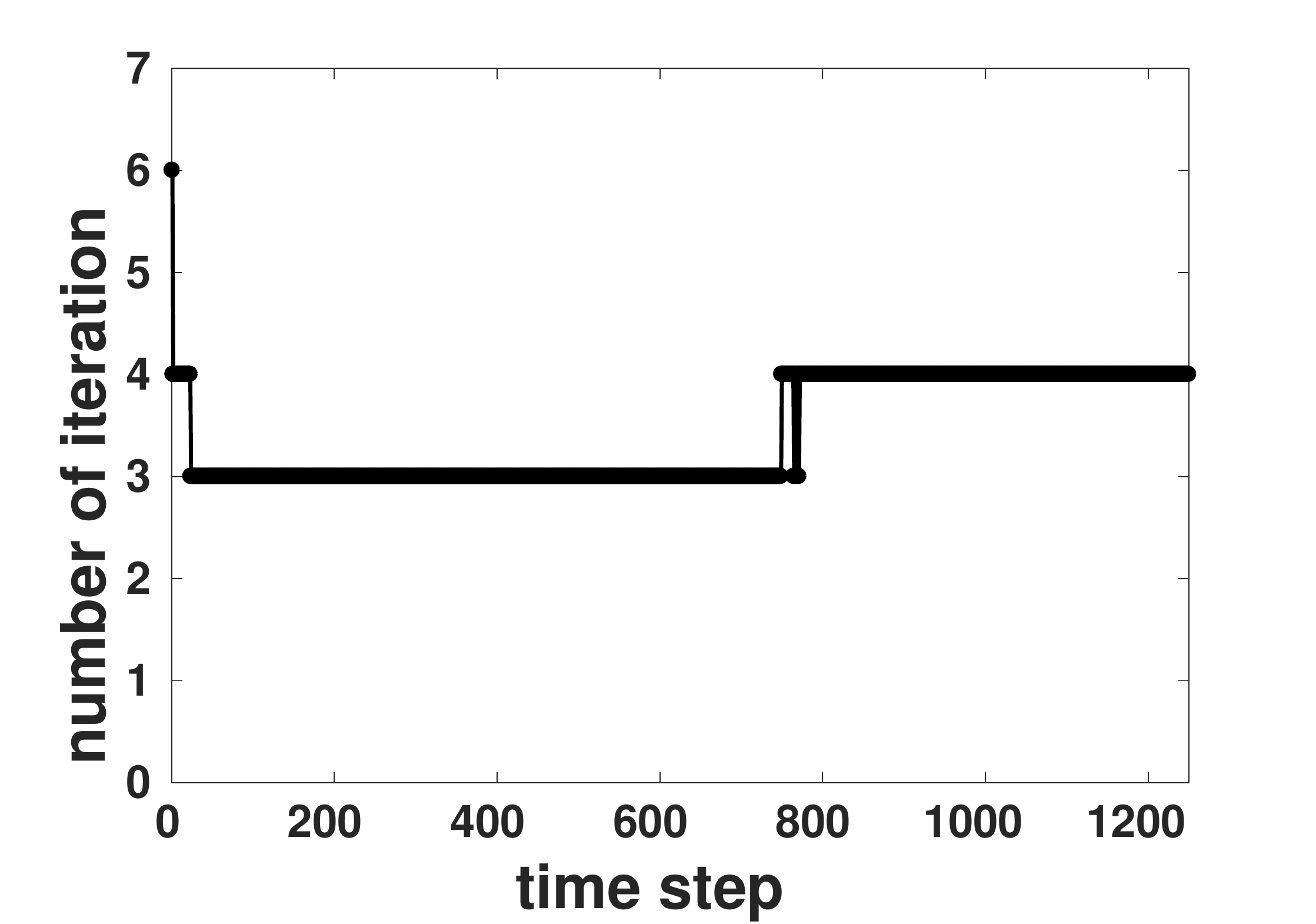}
		\caption{mesh size $256\times 256 \times 320$}
	\end{subfigure}
	\begin{subfigure}[b]{0.32\textwidth}
		\includegraphics[width=\textwidth]{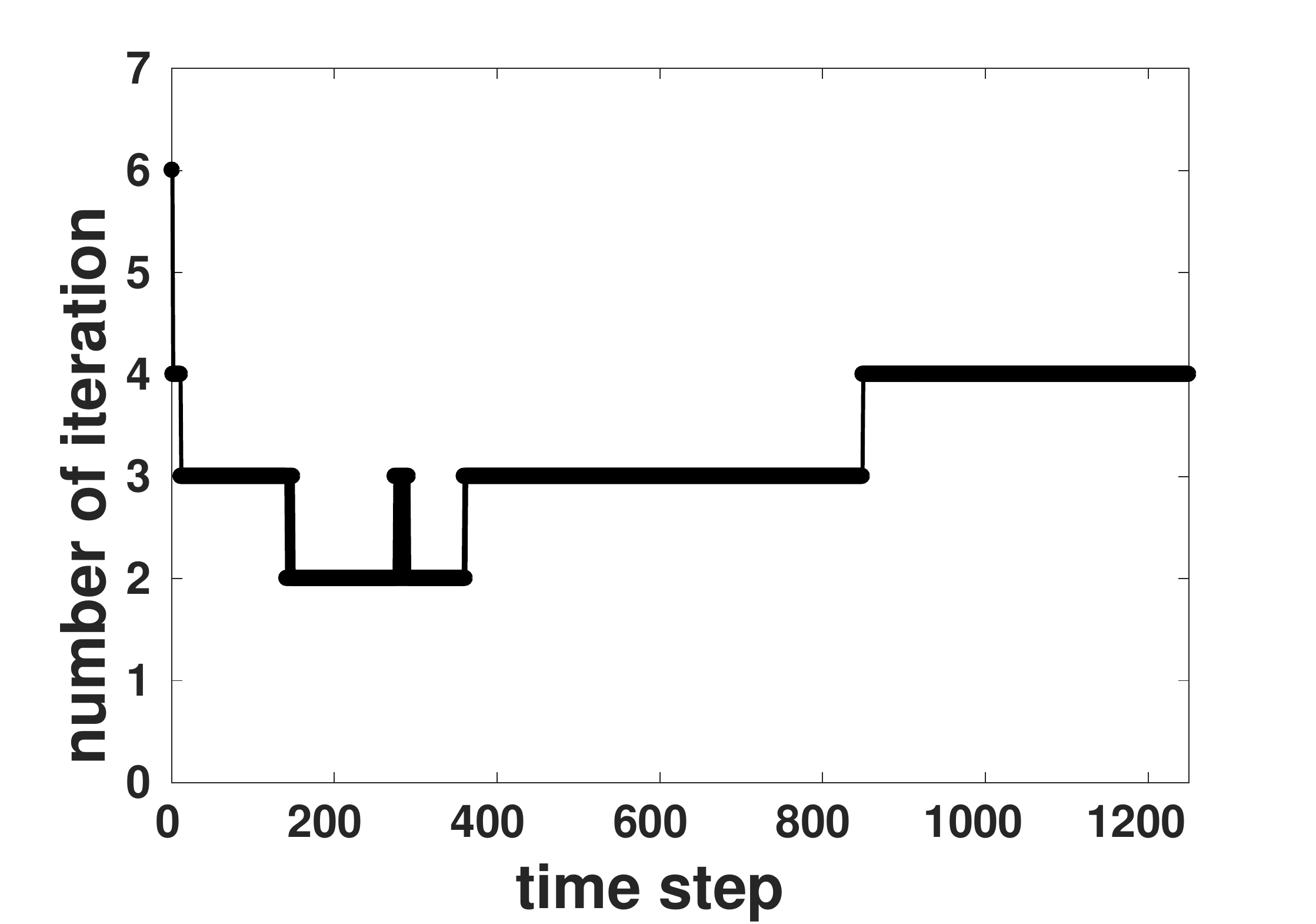}
		\caption{mesh size $512\times 512 \times 640$}
	\end{subfigure}
	\begin{subfigure}[b]{0.32\textwidth}
		\includegraphics[width=\textwidth]{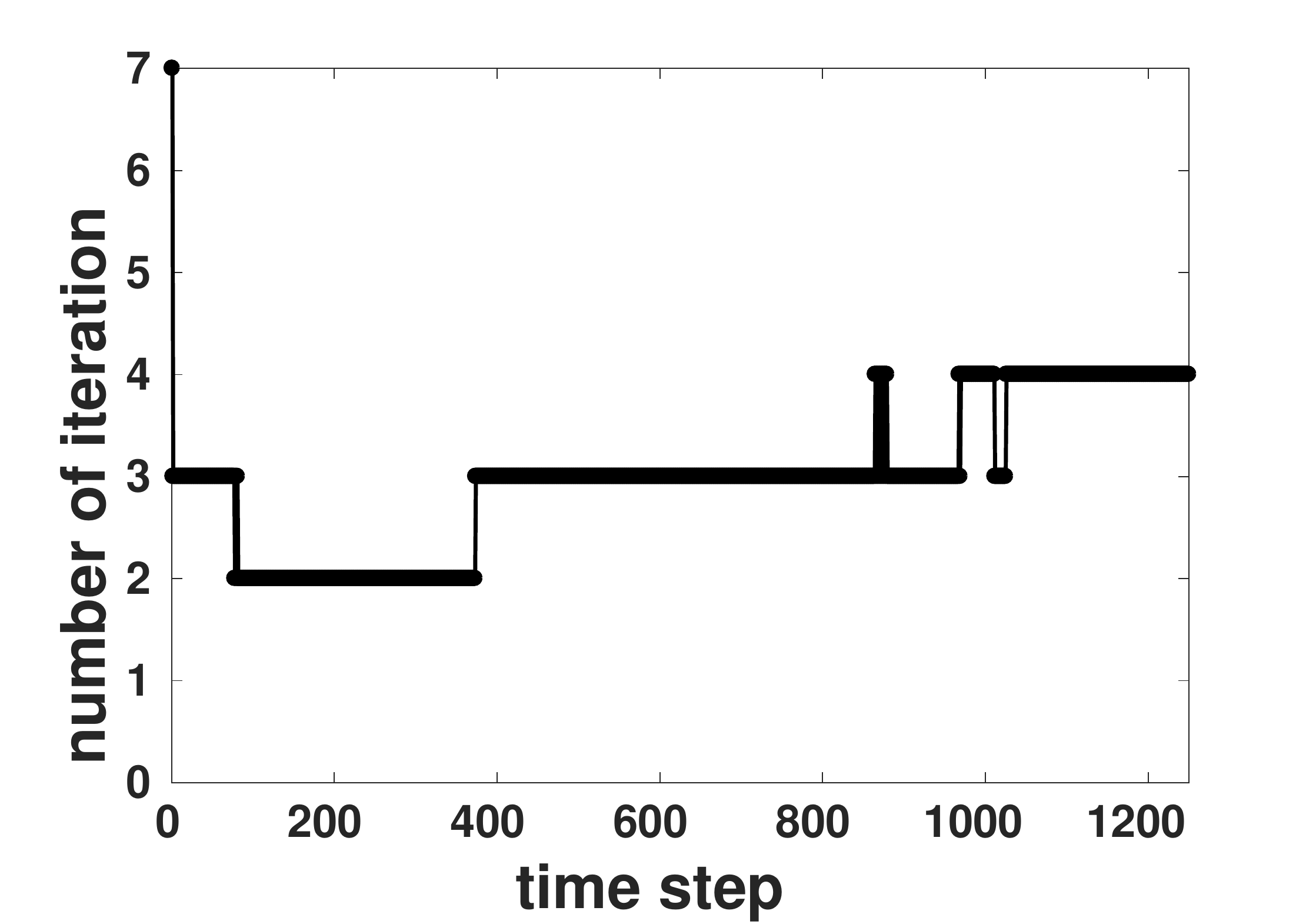}
		\caption{mesh size $1024\times 1024 \times 1280$}
	\end{subfigure}
	\caption{Iteration step for multigrid preconditioned FGMRES solver at each time step by the second-order semi-implicit scheme, on three different meshes.}
	\label{fstep}
\end{figure}

\begin{figure}[htbp]
	\centering
	\begin{subfigure}[b]{0.32\textwidth}
		\includegraphics[width=\textwidth]{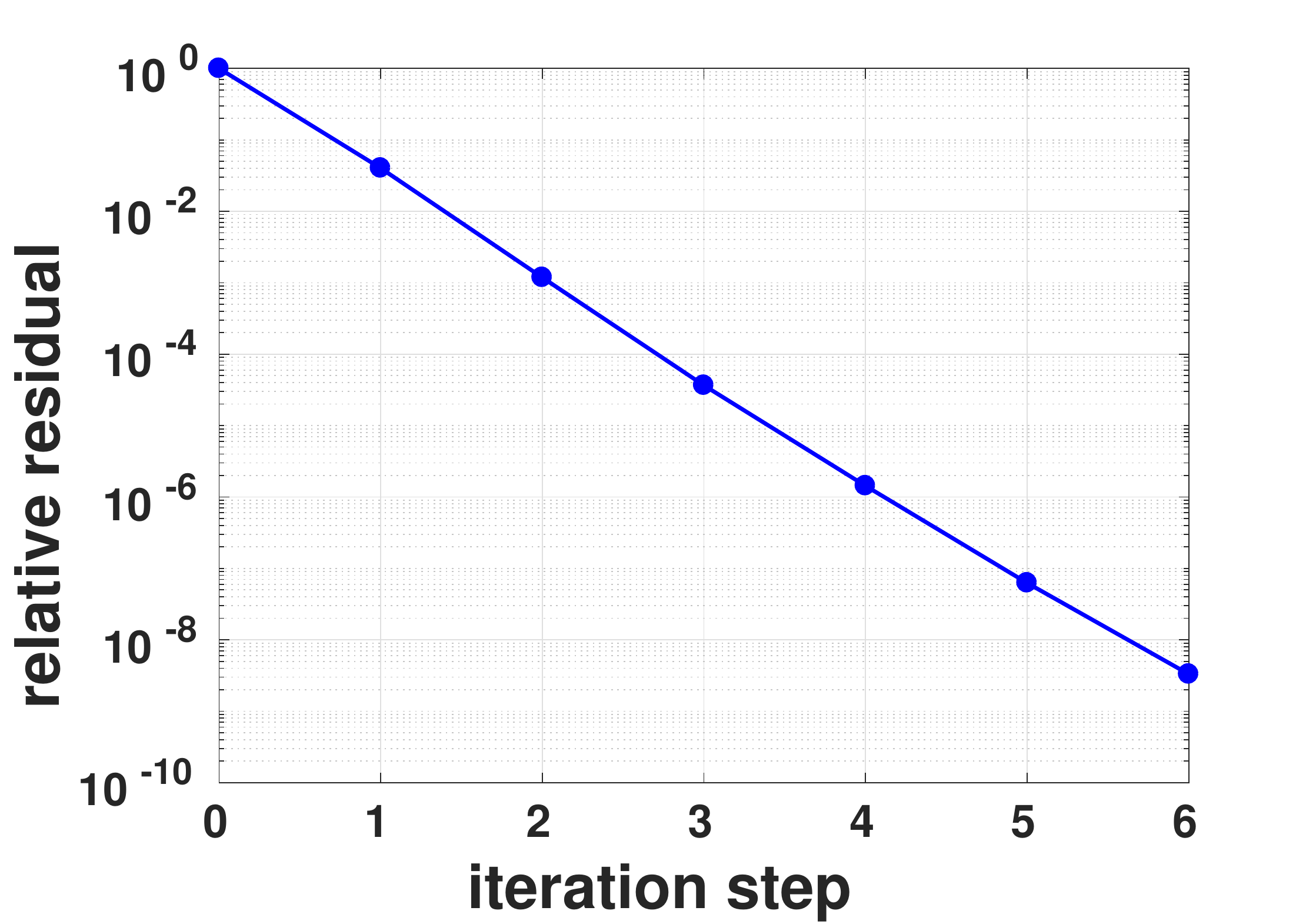}
		\caption{\footnotesize1$^{\text{st}}$ step, mesh $256\times 256 \times 320$}
	\end{subfigure}
	\begin{subfigure}[b]{0.32\textwidth}
		\includegraphics[width=\textwidth]{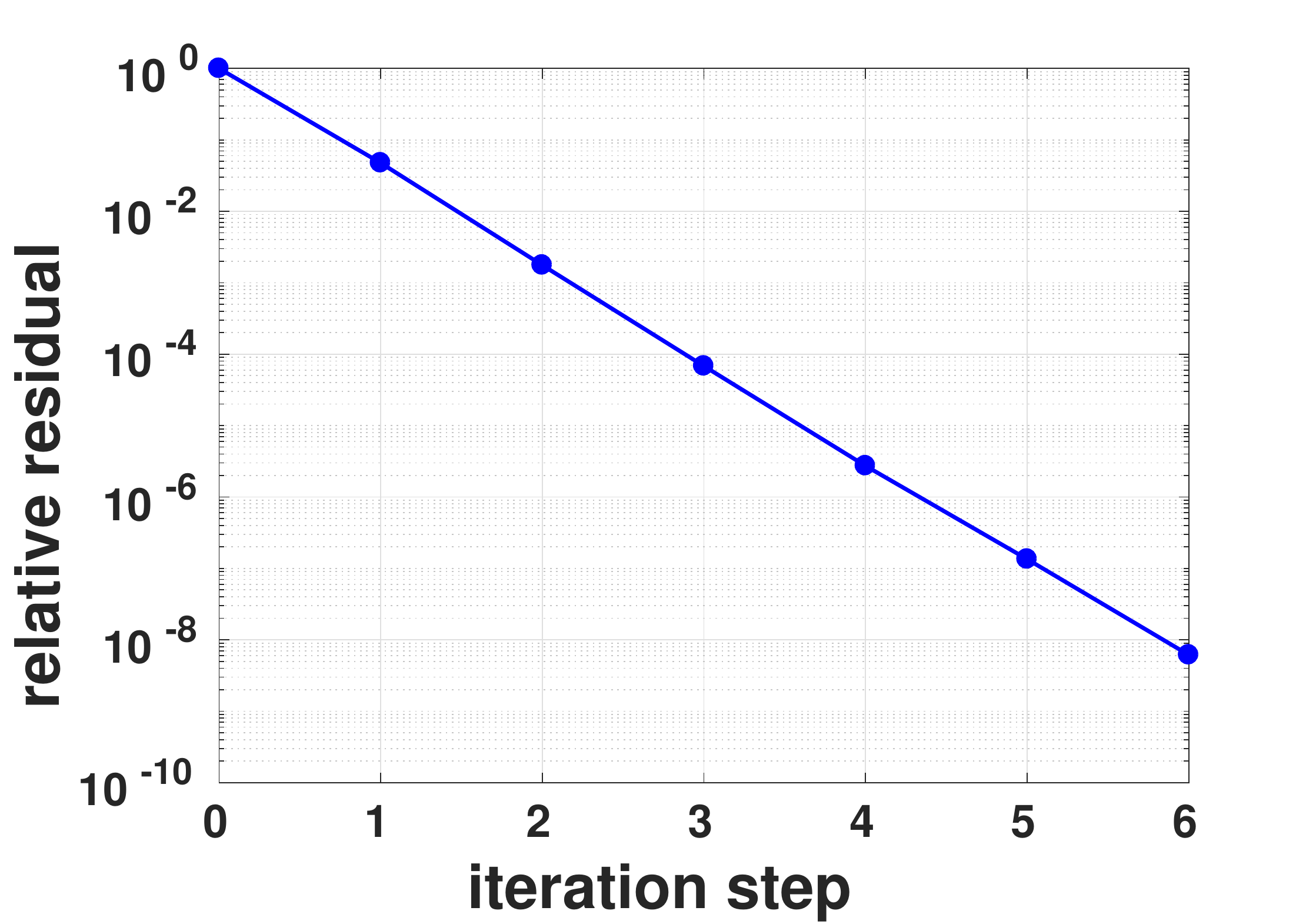}
		\caption{\footnotesize1$^{\text{st}}$ step, mesh $512\times 512 \times 640$}
	\end{subfigure}
	\begin{subfigure}[b]{0.32\textwidth}
		\includegraphics[width=\textwidth]{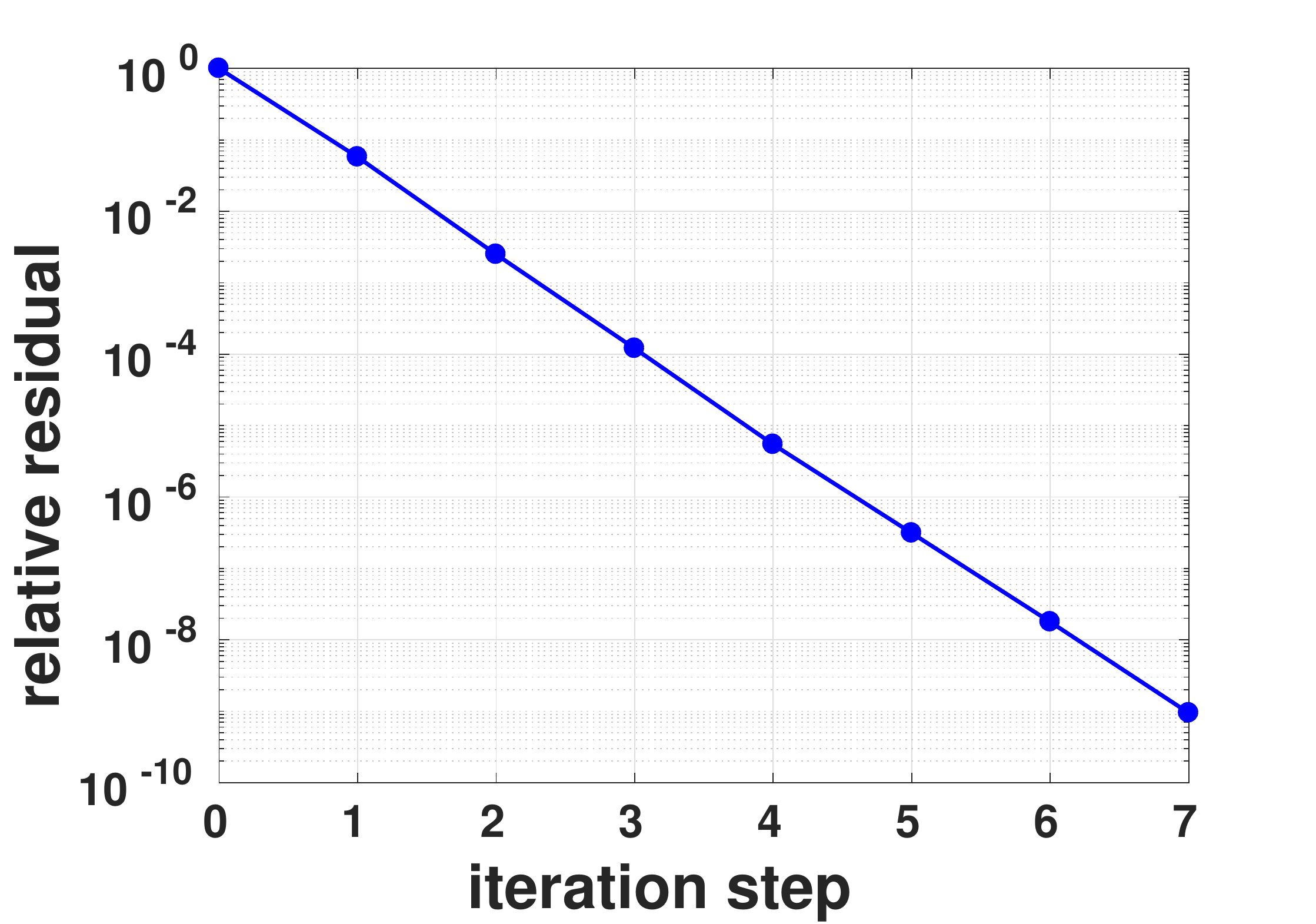}
		\caption{\footnotesize1$^{\text{st}}$ step, mesh $1024\times 1024 \times 1280$}
	\end{subfigure}
	\caption{Relative residual at each iteration for the multigrid preconditioned FGMRES solver in first step on three different meshes.}
	\label{ferror}
\end{figure}

The number of iterations at each time step, for the multigrid preconditioned FGMRES solver in second-order semi-implicit scheme \eqref{e1}--\eqref{e2}, is shown in Figure \ref{fstep}, with the tolerance of the residual is set to $10^{-8}$. Except the first time step, the elliptic solver requires only $2$ to $4$ iterations, and the number of iterations does not increase as the mesh is refined. Figure \ref{ferror} shows the rapid reduction of the relative residual. 

Next, we compare our FGMRES solver with other multigrid preconditioned Krylov subspace methods \cite[Chapter 6--9]{saad2003iterative}, including Conjugate Gradient (CG), Conjugate Gradient Squared (CGS), Bi-CGSTAB (BiCGSTAB), GMRES, and Flexible Conjugate Gradients (FCG). For all these algebraic solvers, we use the same geometric multigrid preconditioner described in Section \ref{multigrid}, with the same convergence criterion $\varepsilon=10^{-8}$. From the previous test, we can see that the number of iteration steps does not vary significantly in the evolutionary process. Hence, for a quick test, we run the same simulation using the same parallel settings but for only 50 steps with a slightly larger time step ($\tau_n = 3.2811$ ps). Table \ref{t1} shows the times consumed by the different Krylov subspace solvers; each time is the average of five runs. Manifestly, FGMRES works best with the multigrid preconditioner, consuming the least computation time in all cases.

\begin{table}[htbp]
\caption{Time costs of different elliptic solvers using 640 cores over 50 time steps using the proposed second-order semi-implicit scheme.}
\centering
\begin{tabular}{l l l l l l l}
\hline
\multicolumn{7}{c}{Mesh size: $256\times256\times320$} \\
\hline
Method & FGMRES & GMRES & CG & FCG & CGS & BiCGSTAB \\
\hline
Mean time [s] & 10.589 & 12.588 & 12.584 & 12.415 & 14.084 & 14.044 \\
Ratio (on FGMRES) & 1 & 1.1888 & 1.1884 & 1.1724 & 1.3301 & 1.3263 \\
\hline
\multicolumn{7}{c}{Mesh size: $512\times512\times640$} \\
\hline
Method & FGMRES & GMRES & CG & FCG & CGS & BiCGSTAB \\
\hline
Mean time [s] & 75.401 & 91.058 & 91.696 & 91.913 & 103.19 & 103.31 \\
Ratio (on FGMRES) & 1 & 1.2076 & 1.2161 & 1.2190 & 1.3685 & 1.3701 \\
\hline
\multicolumn{7}{c}{Mesh size: $1024\times1024\times1280$} \\
\hline
Method & FGMRES & GMRES & CG & FCG & CGS & BiCGSTAB \\
\hline
Mean time [s] & 511.38 & 654.50 & 655.65 & 659.29 & 741.57 & 740.00 \\
Ratio (on FGMRES) & 1 & 1.2799 & 1.2821 & 1.2892 & 1.4501 & 1.4471 \\
\hline
\end{tabular}
\label{t1}
\end{table}

We also implement and test the second-order explicit scheme \eqref{heun_ex_1}--\eqref{heun_ex_3} using the same configurations. The results are shown in Table \ref{t2}, and the FGMRES solver still performs outstandingly. Here we emphasize again that in the explicit scheme, we need to solve the constant coefficient Poisson equation that requires matrix assembly only at the initial step. Even so, for most of the Krylov subspace solvers, the times shown in Table \ref{t2} are larger than the times in Table \ref{t1}. This result occurs because in each time step, our second-order semi-implicit method needs to solve the elliptic problem only once, while the explicit method needs twice, provided that our solver is efficient for both coefficient constant or varied Poisson equation. The only exception is the FGMRES method, in which the explicit scheme has better performance.

\begin{table}[htbp]
	\caption{Time costs of different elliptic solvers using 640 cores over 50 time steps by the second-order explicit scheme.}
	\centering
	\begin{tabular}{l l l l l l l}
		\hline
		\multicolumn{7}{c}{Mesh size: $256\times256\times320$} \\
		\hline
		Method & FGMRES & GMRES & CG & FCG & CGS & BiCGSTAB \\
		\hline
		Mean time [s] & 10.306 & 13.550 & 13.724 & 13.941 & 17.039 & 17.014 \\
		Ratio (on FGMRES) & 1 & 1.3148 & 1.3317 & 1.3527 & 1.6533 & 1.6509 \\
		\hline
		\multicolumn{7}{c}{Mesh size: $512\times512\times640$} \\
		\hline
		Method & FGMRES & GMRES & CG & FCG & CGS & BiCGSTAB \\
		\hline
		Mean time [s] & 61.028 & 94.798 & 95.294 & 95.398 & 119.30 & 119.61 \\
		Ratio (on FGMRES) & 1 & 1.5534 & 1.5615 & 1.5632 & 1.9548 & 1.9599 \\
		\hline
		\multicolumn{7}{c}{Mesh size: $1024\times1024\times1280$} \\
		\hline
		Method & FGMRES & GMRES & CG & FCG & CGS & BiCGSTAB \\
		\hline
		Mean time [s] & 352.88 & 694.12 & 689.98 & 695.97 & 867.26 & 865.01 \\
		Ratio (on FGMRES) & 1 & 1.9670 & 1.9553 & 1.9723 & 2.4577 & 2.4513 \\
		\hline
	\end{tabular}
	\label{t2}
\end{table}

As Tables \ref{t1} and \ref{t2} show, the time consumption is roughly proportional to the number of DOFs, indicating the excellent performance of the multigrid preconditioner. In fact, the mean time for all the solvers scales sublinearly. One possible reason is that the average number of iterative steps for a small grid size may be larger, as shown in Figure \ref{fstep}. In addition, the communication time may grow only sublinearly, especially when the amount of data being transferred is small.

Our results also show that the proposed algebraic elliptic solver outperforms the R\&B SOR solver introduced in \cite{0022-3727-51-9-095206}, which consumes approximately 55 seconds on 200 cores to converge to a relative residual lower than $10^{-6}$ on an $800^3$ grid (see Figure 6(c) in \cite{0022-3727-51-9-095206}). For comparison purposes, we assume a linear speedup for R\&B SOR; then, the same solver costs $55/(640/200) \approx 17.2$ seconds on 640 cores. In our numerical test, even for a much larger grid size $1024 \times 1024 \times 1280$ ($\approx 2.62 \times 800^3$), solving a single linear system to a tolerance of $10^{-8}$ requires only $352.88 / (50 \times 2) = 3.52$ seconds. Here, we use the value from Table \ref{t2} because the constant coefficient Poisson equation was solved in \cite{0022-3727-51-9-095206}. In addition, Figure 6(b) and Figure 6(c) in \cite{0022-3727-51-9-095206} imply that the time complexity of the R\&B SOR method is higher than $O(N)$(even higher than $O(N \log N)$) where $N$ is the total number of DOFs. It should be remarked that the performance depends on the hardware and network, and this comparison is based on the data obtained on two independent high performance clusters. The cluster in \cite{0022-3727-51-9-095206} is built using Intel Xeon E5-2680 v2 processors (10 cores, 2.8 GHz) and that of Tianhe2-JK are Intel Xeon E5-2660 v3 CPUs (10 cores, 2.6 GHz).

\section{Applications}
In this section, we carry out two applications with different initial settings to compare the 3D results with the results obtained by the 2D moving mesh method, and to study the interactions of two streamers. The semi-implicit method \eqref{e1}--\eqref{e2} with adaptive time stepping (\ref{timedrift}) is used. When choosing the time step for our second-order semi-implicit scheme, we are unable to choose $\tau_n$ according to \eqref{timedrift} directly because $\tau_n$ should be given before we solve $\phi^{n+1/2}$ in \eqref{e18}, and $\vec{E}^{n+1/2}$ can be computed only after obtaining $\phi^{n+1/2}$. Therefore, as an alternative, we may choose a relatively small $\tau_0$ at the first step, and then update time step by \eqref{timedrift} with $\vec{E}^{n+1/2}$ is replaced as $\vec{E}^{n-1/2}$ and multiplying a safety factor 0.5 to it. We use simplified conditions without considering detailed models like the chemical reactions, because we aim towards code verification and proof of principle, which is by itself already very challenging for streamer simulations.

\subsection{Double-headed streamer propagation} \label{dh}
The first application is the double-headed streamer in a homogeneous field.
A computational domain $\Omega = (0,1) \times (0,1) \times (0,1)$ cm$^{3}$ is used; however, the numerical results are nearly indistinguishable from the results computed on a larger domain $(0,2) \times (0,2) \times (0,1)$ cm$^{3}$. The initial charge $\tilde{n}(\vec{x})$ is a Gaussian located at the center of the domain (see \eqref{eq51}). The transport coefficients are the same as that are used in Section \ref{scalability} and shown below \eqref{eq51}. 

A very fine mesh with $2048 \times 2048 \times 2560$ cells is used in the simulation, comprising a total of more than 10.7 billion cells. For MPI parallelism, 1280 CPU cores (64 nodes) are used. The initial time step is set to $2$ ps, and the adaptive time step is subsequently selected. The simulation requires 1558 steps to reach the final time 2.5 ns, resulting in an average time step of 1.6 ps. 

\begin{figure}[htbp]
	\centering
	\includegraphics[width=0.7\textwidth]{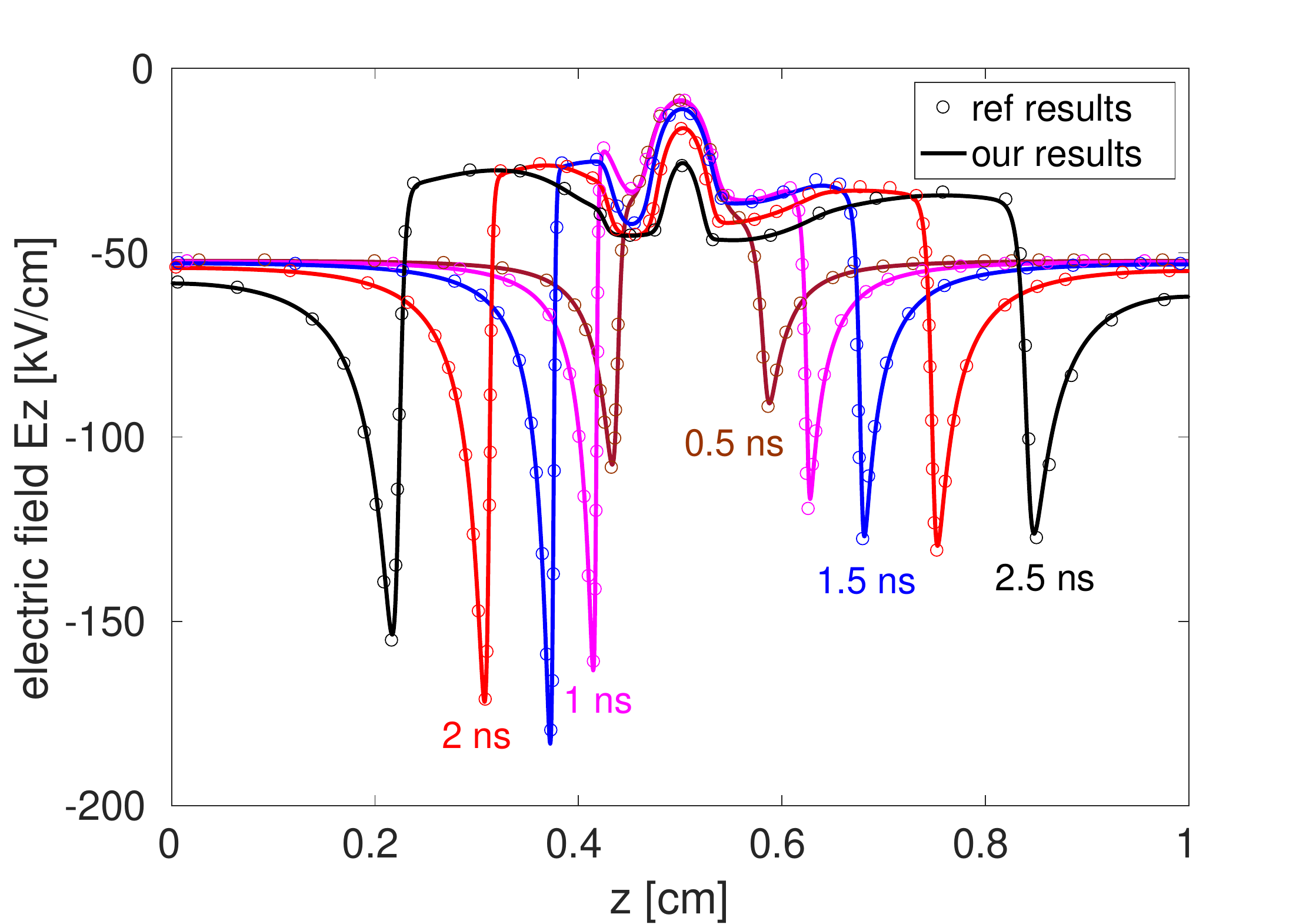}
	\caption{Electric field $E_z$ along the line $x=0.5$, $y=0.5$ in a double-headed streamer (reference data taken from \cite{movingmesh2007}).}
	\label{f2}
\end{figure}

Figure \ref{f2} shows the electric field along the $z$-axis at the center of the streamer channel ($x=y=0.5$). The result in \cite{movingmesh2007}, which used moving mesh method in 2D simulation, is provided as a reference, and the two are in close agreement. At $t=2$ ns, the positive (cathode-directed) and the negative (anode-directed) streamer move approximately 0.19 and 0.26 cm,
respectively, measured using the position of the highest electric field strength. At $t=2.5$ ns, they move approximately 0.28 and 0.36 cm. Hence, the average velocities of negative and positive streamer from 2 to 2.5 ns are estimated approximately $2.0\times10^8$ and $1.95\times10^8$ cm/s, in other words, no obvious difference. Although the negative streamer propagates faster than the positive
streamer at the beginning of the propagation, the difference in velocity becomes much smaller after 1.5 ns. The experimentally measured velocities of streamers vary under different experimental conditions by an order of magnitude \cite{Zeng_2013}, but the typical measured velocities are similar to those obtained in this simulation.

\begin{figure}[htbp]
	\centering
	\begin{subfigure}[b]{0.45\textwidth}
	\includegraphics[width=\textwidth]{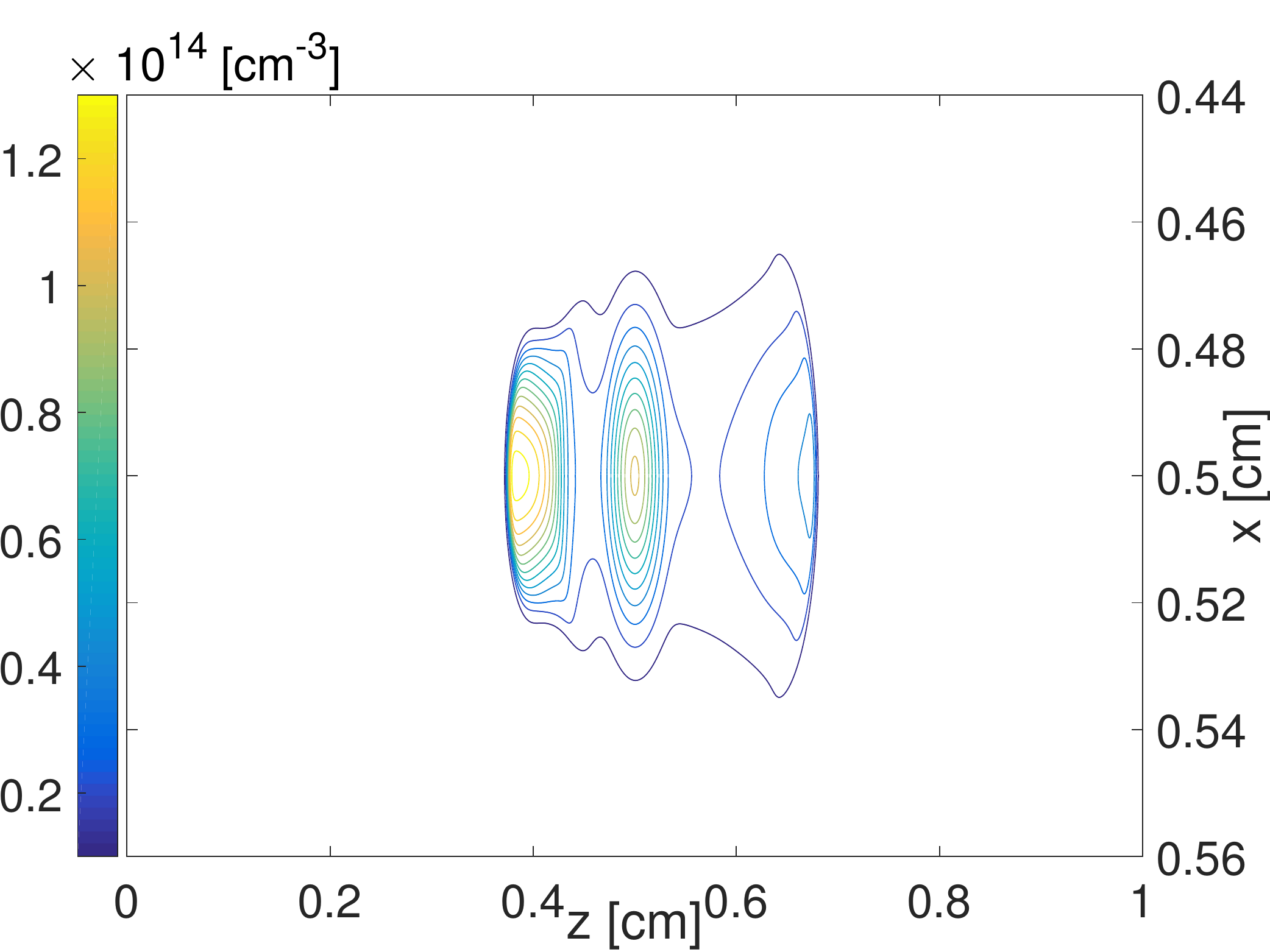}
	\caption{1.5 ns}
	\end{subfigure}
	\hspace{0.05\textwidth}
	\begin{subfigure}[b]{0.45\textwidth}
	\includegraphics[width=\textwidth]{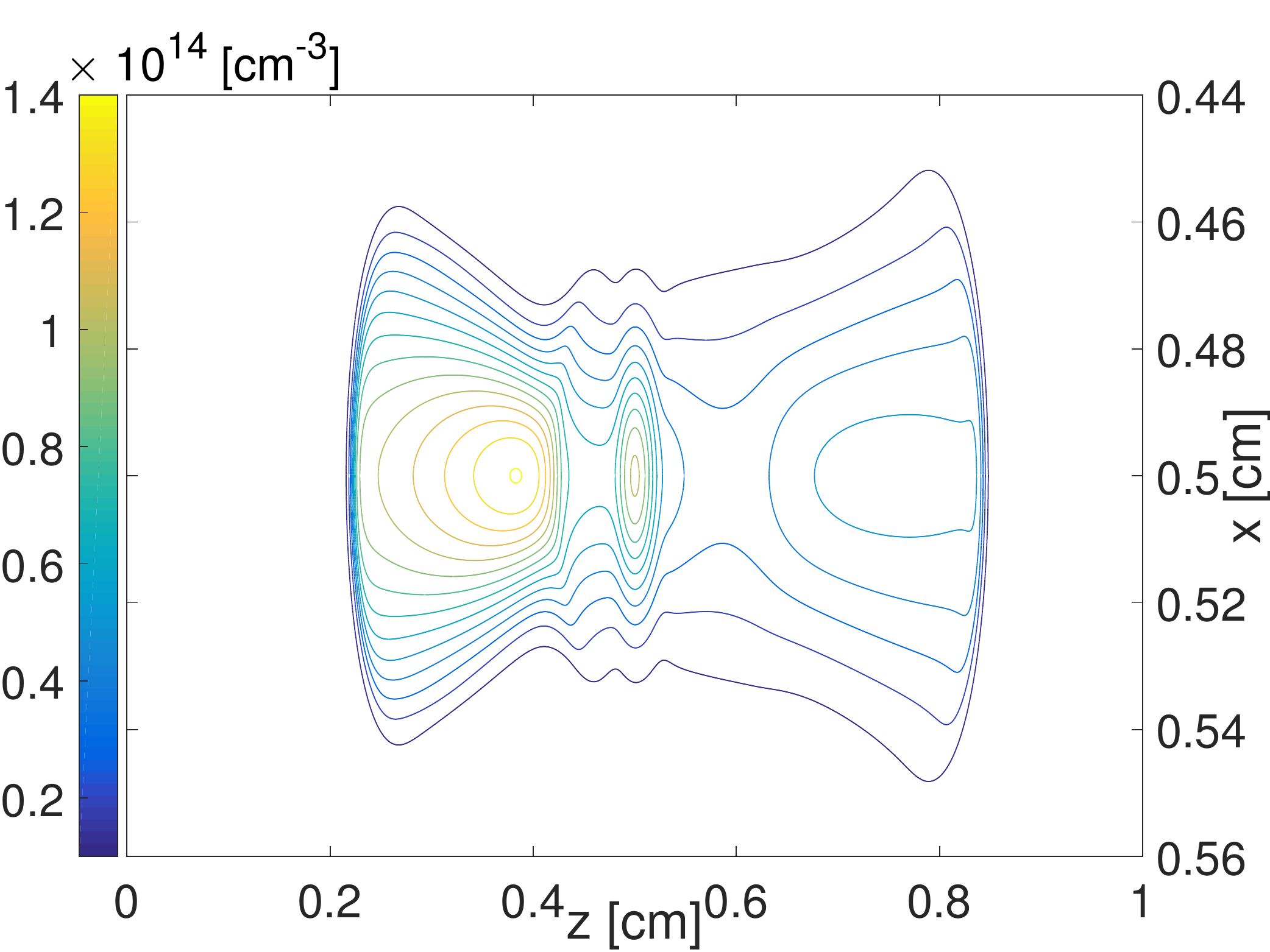}
	\caption{2.5 ns}
	\end{subfigure}
	\caption{Electron density on the $y=0.5$ plane in a double-headed streamer at 1.5 and 2.5 ns.}
	\label{f1}
\end{figure}

\begin{figure}[htbp]
	\centering
	\begin{subfigure}[b]{0.45\textwidth}
		\includegraphics[width=\textwidth]{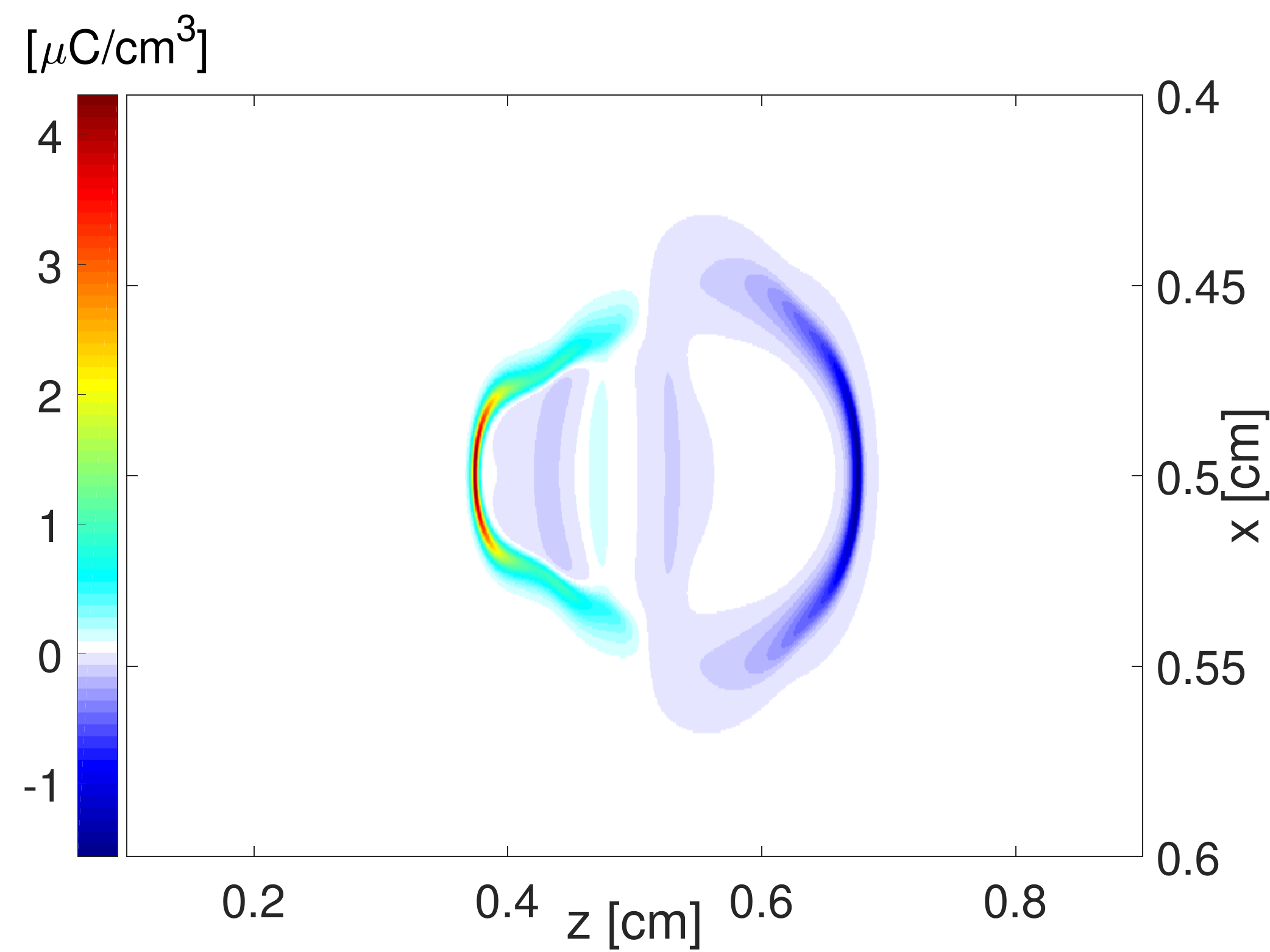}
		\caption{1.5 ns}
	\end{subfigure}
	\hspace{0.05\textwidth}
	\begin{subfigure}[b]{0.45\textwidth}
		\includegraphics[width=\textwidth]{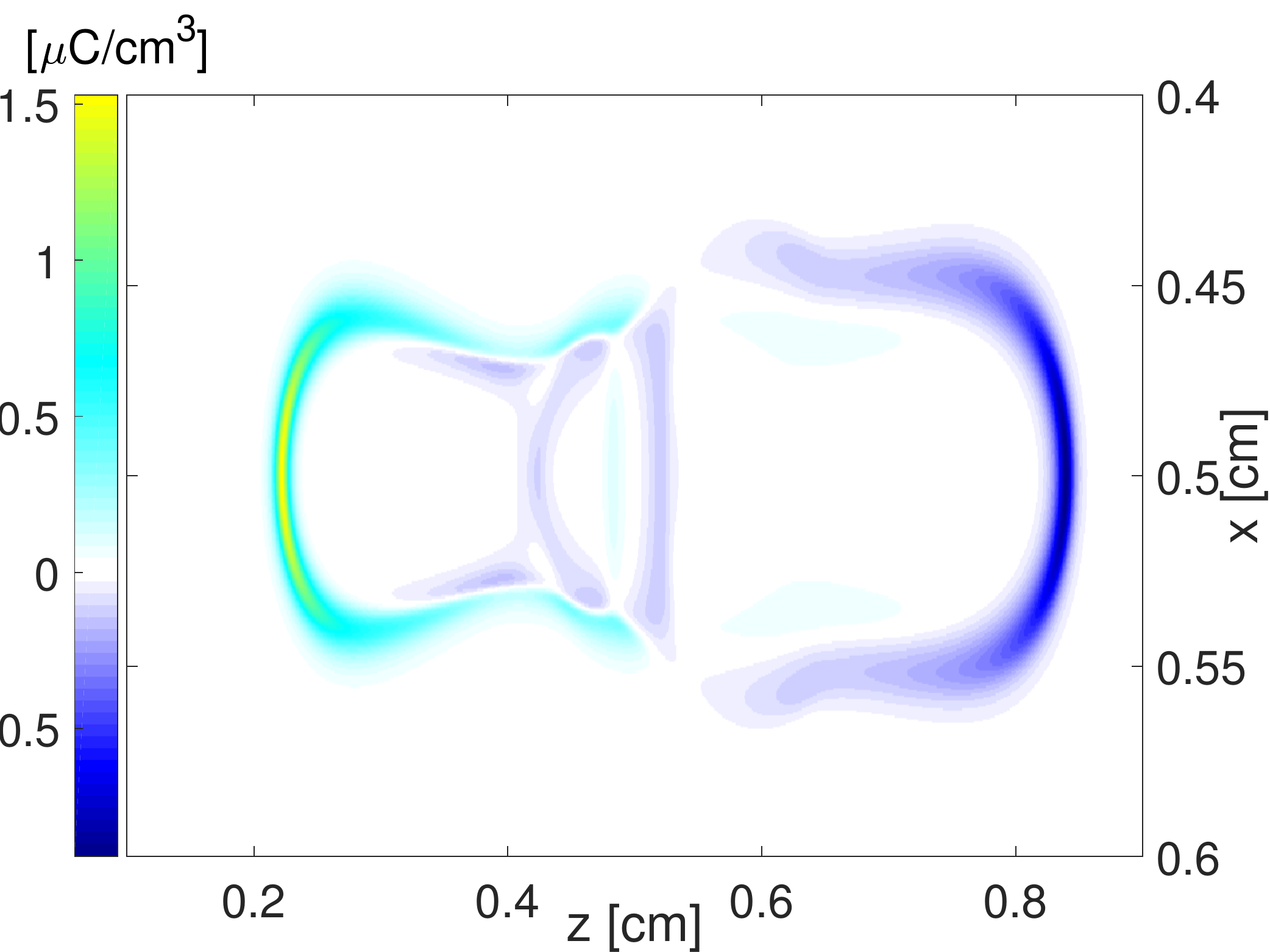}
		\caption{2.5 ns}
	\end{subfigure}
        \caption{Net charge density (in terms of charge per unit volume) on the $y=0.5$ planes in a double-headed streamer at 1.5 and 2.5 ns. Here net charge density equals to the density of the charge carriers ($n_p-n_e$) times the elementary charge ($1.602\times10^{-13}\mu$C). } 
	\label{f3}
\end{figure}

The contours of the electron density and net charge densities on the plane $y = 0.5$ are shown in Figure \ref{f1} and Figure \ref{f3}, respectively. The results in Figure \ref{f1} are very similar to the results obtained in \cite{movingmesh2007}. Figure \ref{f3} clearly shows that at the front of the streamers' head, a thin layer of net charge that has the same polarity as the streamer exists. The thickness of this layer is approximately 0.2 mm to 0.3 mm, and the maximum net charge density is on the order of 1$\mu$C/cm$^3$, which is approximately between $10^{12}$ and $10^{13}$ charged particles per cm$^3$. 

\subsection{Interactions of two cathode-directed streamers}
The second application considers the interactions of two cathode-directed streamers. A 3D simulation enables study of streamer discharges without assuming the axisymmetry. The settings of these simulations are the same as those described in Section \ref{dh} and \ref{scalability}, except that the initial conditions are set to be the sum of two Gaussians:
\begin{align}
\begin{aligned}
\tilde{n}(\vec{x}) = 10^8 + 10^{14} \Big[& \exp \left( - \left( \frac{(x-0.5-x_0)^2 + (y-0.5)^2+(z-1)^2}{\sigma^2} \right) \right) \\ 
+ & \exp \left( - \left( \frac{(x-0.5+x_0)^2 + (y-0.5)^2+(z-1)^2}{\sigma^2} \right) \right) \Big] ~ \mbox{cm}^{-3}, 
\end{aligned}
\end{align}
where $2x_0$ describes the distance between the centers of the two Gaussian-shaped seed charges, and $\sigma=0.03$. We set $x_0$ to $\sigma$ and $3\sigma$, to study the effects of different distances between the two streamers on the interactions. 

\begin{figure}[htbp]
	\centering
	\begin{subfigure}[b]{0.9\textwidth}
		\centering
		\includegraphics[width=0.4\textwidth]{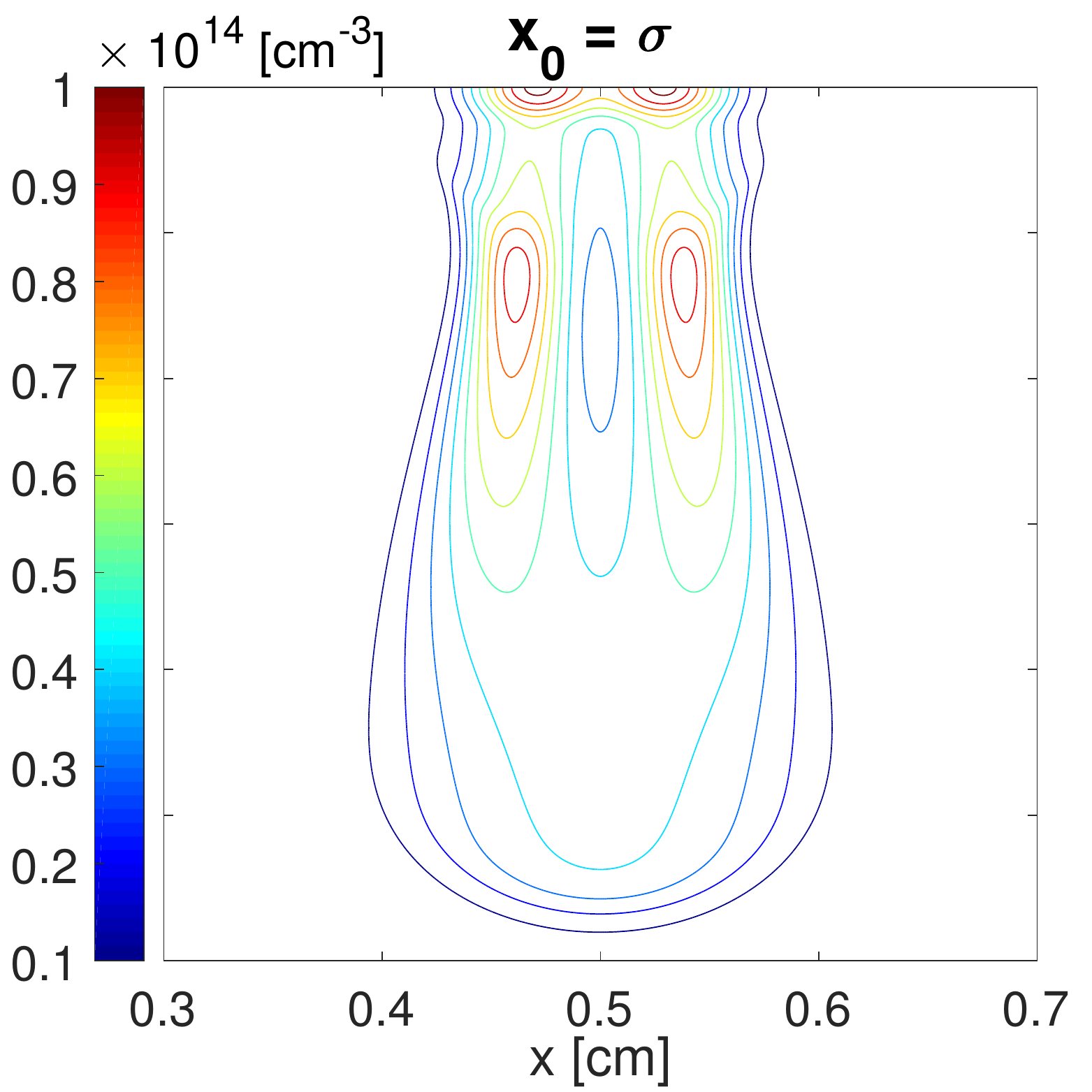}
		\hspace{0.05\textwidth}
		\includegraphics[width=0.4\textwidth]{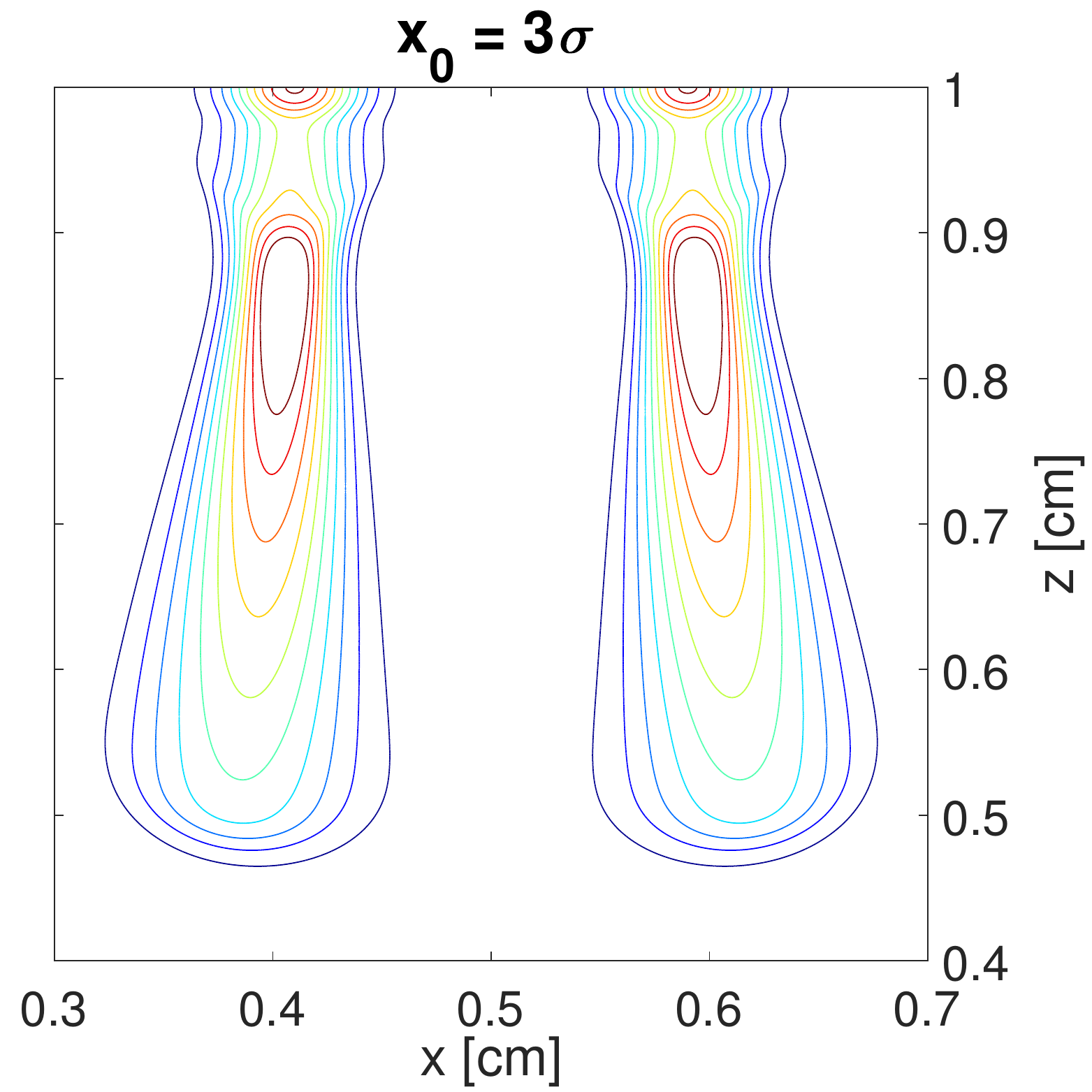}
		\caption{contours of electron density on $y=0.5$ plane}
	\end{subfigure}
	\begin{subfigure}[b]{0.9\textwidth}
		\centering
		\includegraphics[width=0.4\textwidth,bb=400bp 0bp 1200bp 750bp,clip]{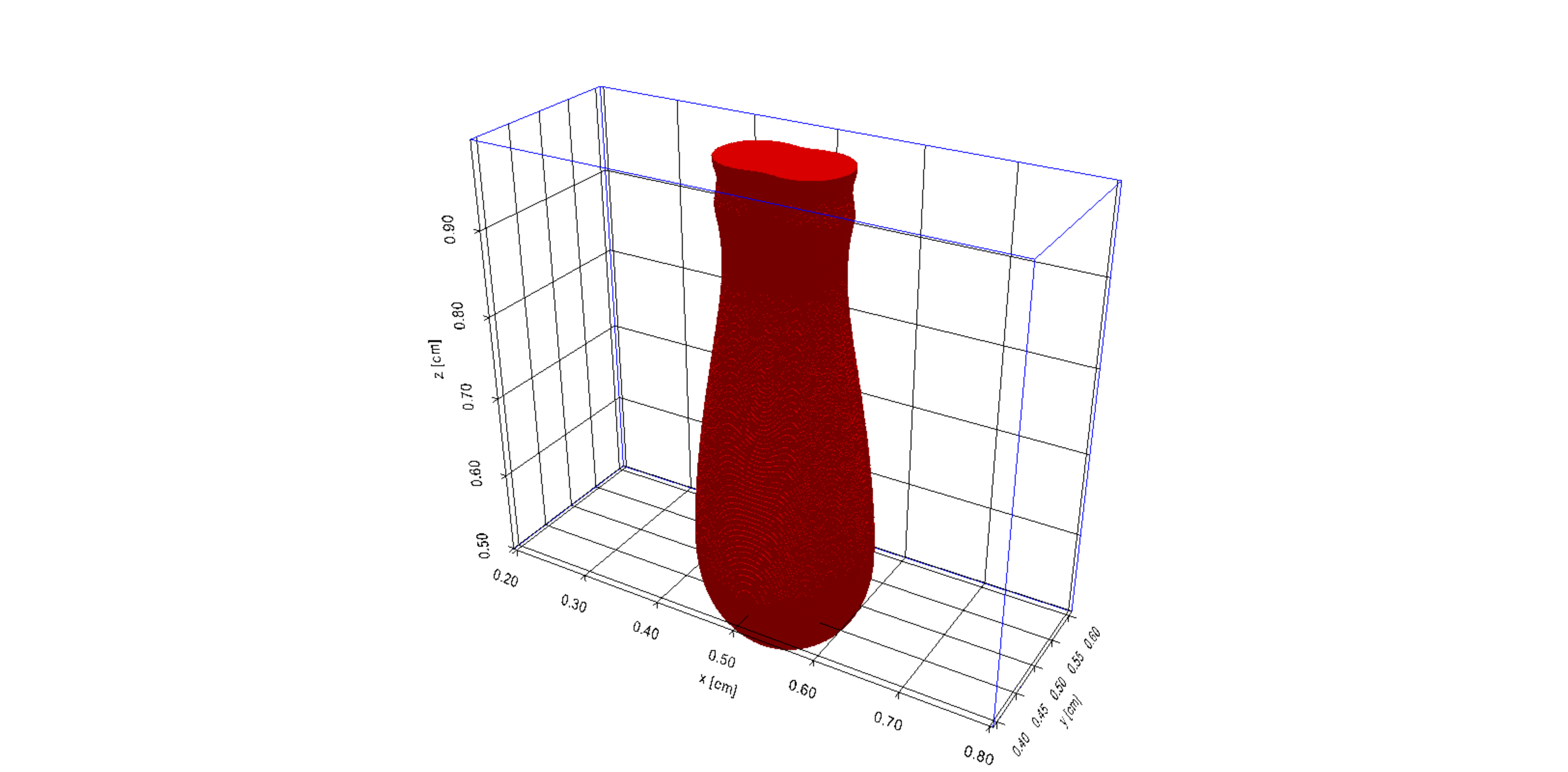}
		\hspace{0.05\textwidth}
		\includegraphics[width=0.4\textwidth,bb=400bp 0bp 1200bp 750bp,clip]{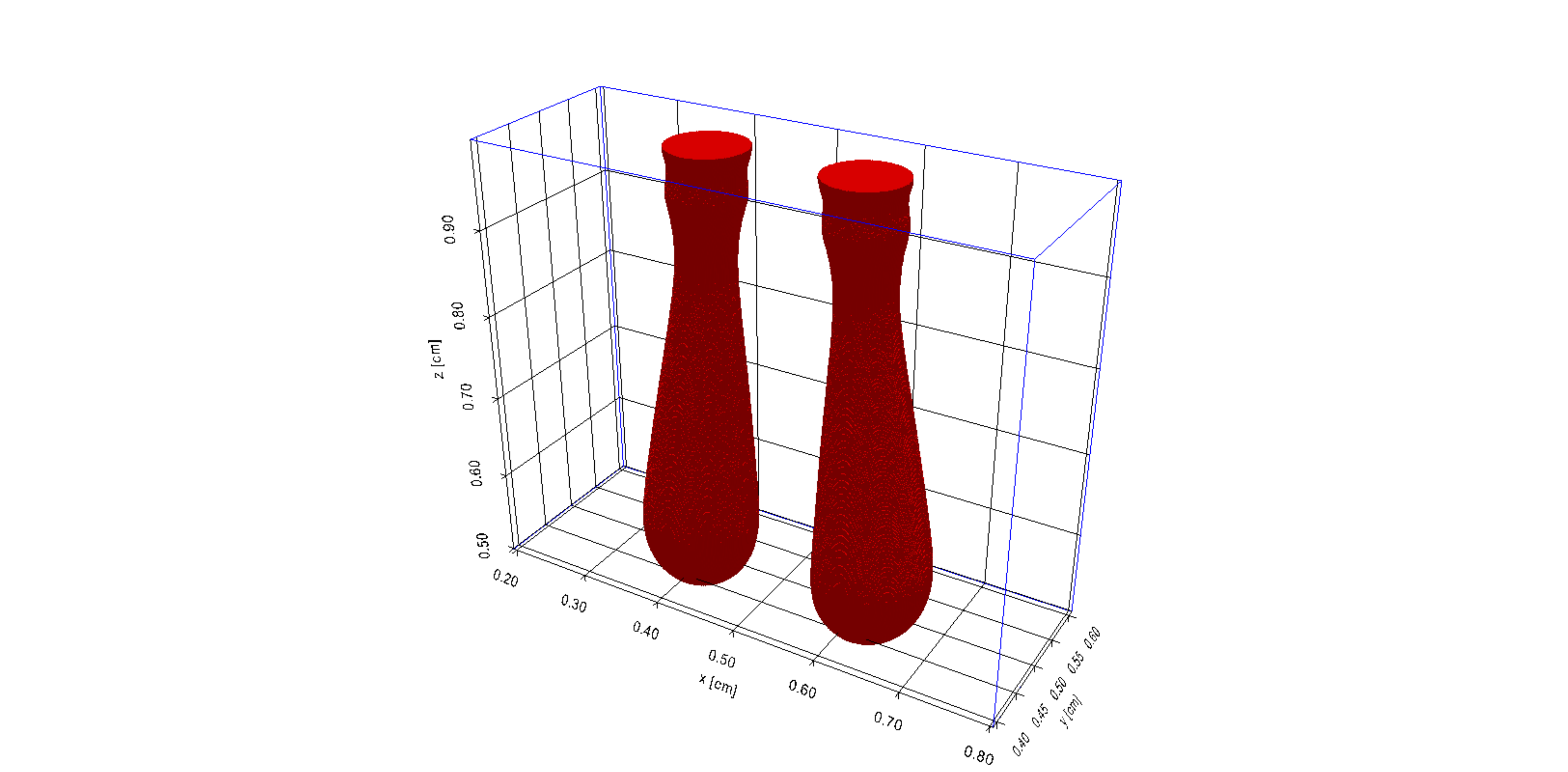}
		\caption{3D plots of electron density greater than 10$^{13}$ cm$^{-3}$}
	\end{subfigure}
	\begin{subfigure}[b]{0.9\textwidth}
		\centering
		\includegraphics[width=0.4\textwidth]{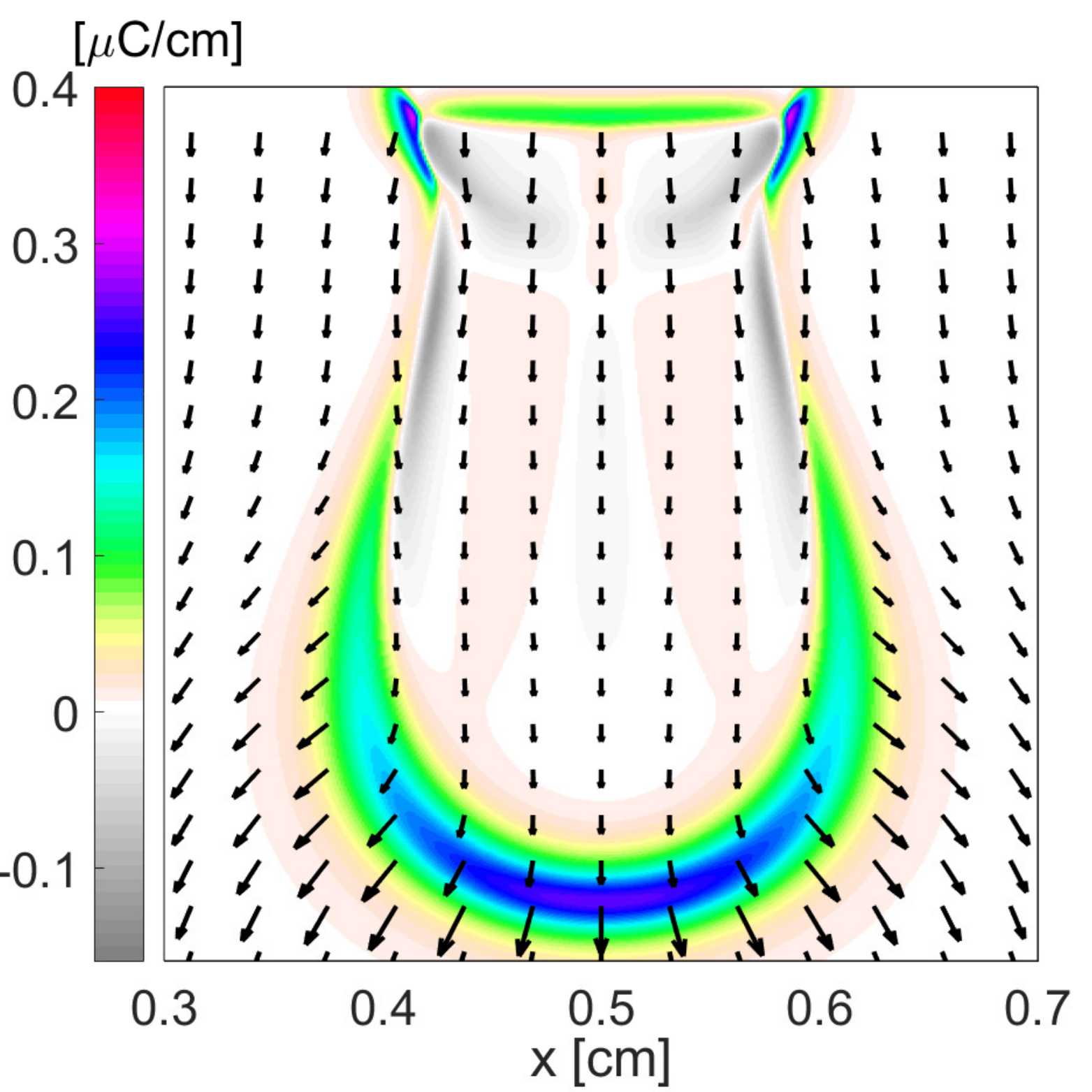}
		\hspace{0.05\textwidth}
		\includegraphics[width=0.4\textwidth]{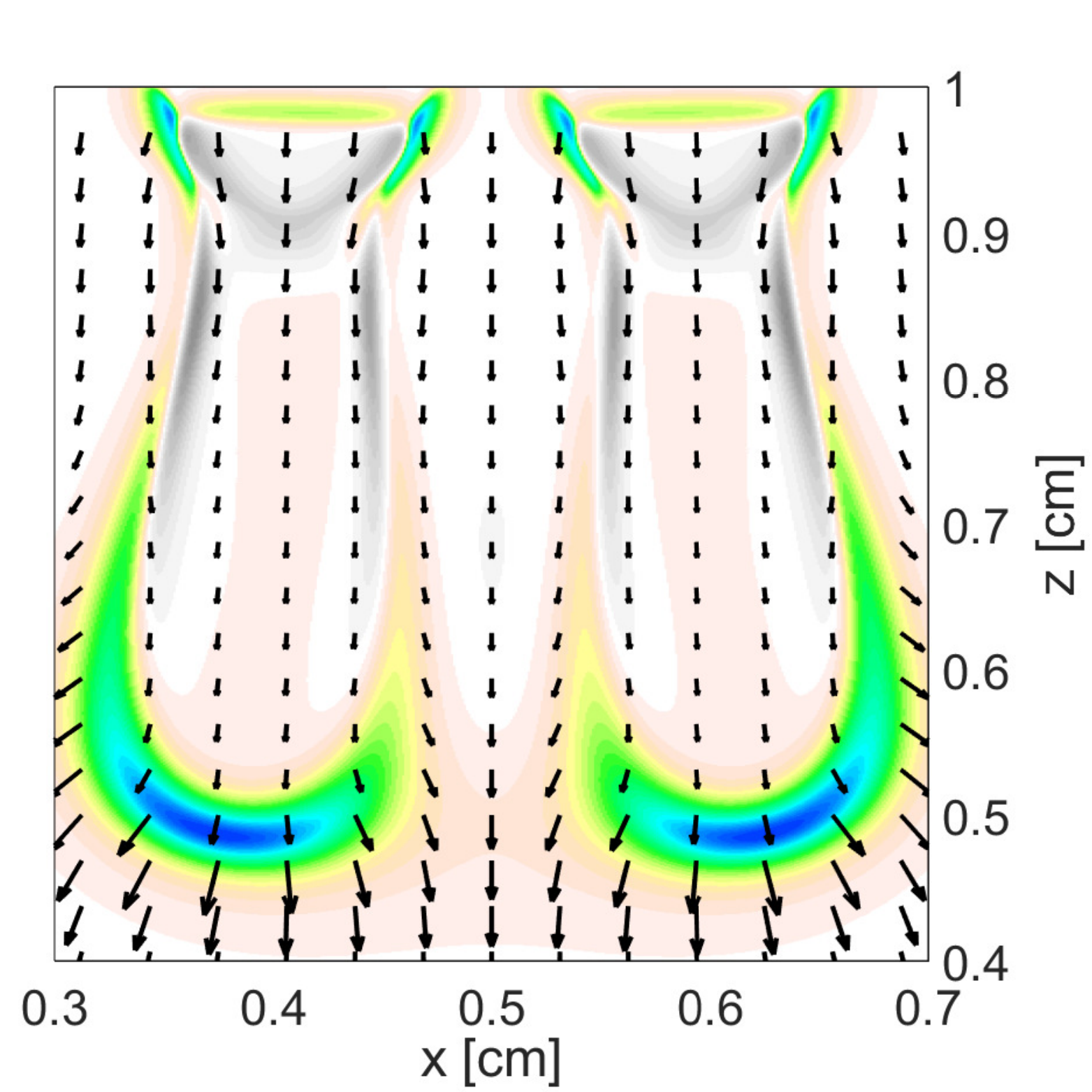}
		\caption{net charge density and electric field $(E_x, E_z)$ on $y=0.5$ plane}
	\end{subfigure}
	\caption{Electron density, net charge density and electric field $(E_x, E_z)$ for the interaction between two streamers when time $t=3.5$\,ns, for $x_0=\sigma$ (left column) and $x_0=3\sigma$ (right column). }
	\label{finteract}
\end{figure}

Figure \ref{finteract} show the evolution of the electric field, net charge and electron density when time $t=3.5$\,ns. 
When the two Gaussians are close (e.g., $x_0=\sigma$), the two streamers clearly merge into one. When the distance between the two Gaussians is slightly larger (e.g., $x_0=3\sigma$), the strong repulsion resulting from the net charge layer at the fronts of the streamers causes the streamers  no longer propagating in the direction of the applied field; however, the distribution of positive ions has already merged by $t=3.5$ ns, and the two streamers become closer as they propagate. 

This may be partly explained as follows. The electric field in the centre of the two streamers (i.e., at $x=0.5$ cm) is along the direction of the applied field and perpendicular to the electrodes, while the electric fields on the left and right sides of this line have opposite directions, both pointing away from the streamers, which drives the electrons toward the streamers and leaves behind the positive ions in this area. However, this positive net charge greatly enhances the local electric field and attracts the seed electrons (which may be generated by nonlocal photoionization \cite{luque}) ahead of it to this area. Hence, the electron density gradually increases due to the collision ionization, which causes the possible merging of the two streamers.

To more focus on the simulator, we here have used a background photoionization rate for simplicity \cite{dhali1987two}. However, the basic observations are consistent with those in \cite{luque} which indicates that two adjacent streamers can interact through electrostatic repulsion and through attraction due to nonlocal photoionization. We will study this using a more detailed photoionization model in future.

\section{Conclusion}
In this paper, we have proposed a simulator using a second-order semi-implicit scheme with multigrid preconditioned FGMRES elliptic solver for 3-dimensional streamer discharge simulations under MPI, which contributes in three main aspects. First, the semi-implicit temporal discretization achieves second-order accuracy; however, it requires solving an elliptic equation only once during each time step. Compared with the explicit schemes, this scheme has better stability because it allows the use of time steps larger than the dielectric relaxation time. Second, we adopt a geometric multigrid preconditioned FGMRES solver to solve the variable coefficient elliptic equation. Throughout our simulations, the elliptic solver requires a small number of iterations (typically 3 to 4) to converge to a relative tolerance of $10^{-8}$. It is numerically verified that FGMRES is faster than R\&B SOR and five other Krylov subspace solvers. Third, our simulations show that our MPI implementation achieves high parallel efficiency and significantly reduces the computation time when the number of cores increases. We conducted two simulations to study the propagation of a double-headed streamer and the interactions between two streamers using more than 10.7 billion grid cells. 

Our future work includes a generalization to curved boundaries, and the use of adaptive mesh strategies which will further accelerate the computation.

\section*{Acknowledgments}
This paper is partly supported by the National Science Foundation of China under Grant 51577098 (C. Zhuang), National University of Singapore Startup Fund under Grant No. R-146-000-241-133 (Z. Cai) and the Academic Research Fund of Ministry of Education of Singapore under Grant No. R-146-000-247-114 (W. Bao). Beijing Computational Science Research Center (CSRC) is acknowledged for providing the Tianhe2-JK computing resource. The first author is grateful to Dr. Yongyong Cai and Dr. Lizhen Chen for their help in CSRC.

\appendix

\section{Explicit multigrid preconditioned matrix}\label{appendixa}

Following the same notation in Section \ref{multigrid}, one step of the V-cycle geometric multigrid solver with two layers in Figure \ref{multigriddiagram} can be expressed as follows:
\begin{align}
\begin{aligned}
x_2^{(3)} 
& = Q_2 [ Q_2 x_2^{(0)}+ P_2 b_2 + P_{1,2}A_1^{-1}R_{2,1}(b_2-A_2 Q_2 x_2^{(0)} - A_2 P_2 b_2) ] + P_2 b_2,
\end{aligned}
\label{e24}
\end{align}
which represents the iteration between $x_2^{(0)}$ and $x_2^{(3)}$. Simplifying \eqref{e24} we obtain
\begin{align}
x_2^{(3)} = [Q_2^{2} - Q_2 P_{1,2}A_1^{-1}R_{2,1}A_2 Q_2] x_2^{(0)} + [Q_2P_2 + P_2 + Q_2 P_{1,2}A_1^{-1}R_{2,1}(I_2-A_2P_2)]b_2.
\label{e311}
\end{align}
Then we denote $M_2=Q_2P_2 + P_2 + Q_2 P_{1,2}A_1^{-1}R_{2,1}(I_2-A_2P_2)$. Note that here, $M_2$ is the same as the one defined in \eqref{eRichard}. The matrix in front of $x_2^{(0)}$ in \eqref{e311} is 
\begin{align}
\begin{aligned}
Q_2^{2} - Q_2 P_{1,2}A_1^{-1}R_{2,1}A_2 Q_2 & = I_2 + Q_2^{2} -I_2 - Q_2 P_{1,2}A_1^{-1}R_{2,1}A_2 Q_2 \\
& = I_2 + (Q_2 + I_2)(Q_2 - I_2) - Q_2 P_{1,2}A_1^{-1}R_{2,1}A_2 Q_2 \\
& = I_2 + (Q_2 + I_2) (-P_2 A_2) + Q_2 P_{1,2}A_1^{-1}R_{2,1}A_2(P_2 A_2 - I_2),
\end{aligned}
\label{e25}
\end{align}
where $Q_2 = I_2 -P_2A_2$ is used in the last line of \eqref{e25} by denoting of $Q_2 = q_{n+1}(A) = (I_2-p_n(A)A)$ and $P_2 = p_n(A)$. Then, 
\begin{align}
\begin{aligned}
& I_2 + (Q_2 + I_2) (-P_2 A_2) + Q_2 P_{1,2}A_1^{-1}R_{2,1}A_2(P_2 A_2 - I_2) \\
=&  I_2 - (Q_2 P_2+ P_2) A_2 + Q_2 P_{1,2}A_1^{-1}R_{2,1}(A_2 P_2 - I_2 )A_2 \\
= & I_2 - [(Q_2 P_2+ P_2) + Q_2 P_{1,2}A_1^{-1}R_{2,1}( I_2 - A_2 P_2 )]A_2 = I_2 - M_2 A_2.
\end{aligned}
\label{e27}
\end{align}
Now it is clear that the V-cycle geometric multigrid method \eqref{e311} is 
\begin{align}
x_2^{(3)} = [I_2 - M_2 A_2] x_2^{(0)} + M_2 b_2 = x_2^{(0)} + M_2 (b_2 - A_2 x_2^{(0)}),
\label{a6}
\end{align}
which is a Richardson iteration with the multigrid preconditioner $M_2$. \eqref{a6} shows \eqref{eRichard}. If more than one layer is taken in a multigrid, we can obtain the preconditioner matrix recursively. The only difference is that we use the multigrid again to obtain the solution on the coarse mesh rather than the inverse of the matrix. Therefore, the multigrid preconditioner matrix $M_l$ for $l$ layers V-cycle multigrid can be expressed recursively as follows:
\begin{align}
\begin{aligned}
M_1 & = A_1^{-1},\\
M_l & = Q_l P_l + P_l + Q_l P_{l-1,l} M_{l-1} R_{l,l-1}(I_l-A_l P_l), \qquad (l \geq 2).
\end{aligned}
\end{align}

\bibliography{ssnalbib}{}
\bibliographystyle{siam}

\end{document}